\documentclass[12pt]{article}
\usepackage{graphicx,axodraw,latexsym}

  \def\pd{\partial} \def\pp{\prime} \def\a{\alpha} \def\b{\beta} \def\dl{\delta} \def\s{\sigma}  \def\vphi{\varphi} \def\eps{\epsilon} 
 \def\lam{\lambda} \def\Lam{\Lambda} \def\gm{\gamma} \def\Gm{\Gamma}
\def\om{\omega} \def\Om{\Omega} \def\nb{\nabla} \def\sq{\sqrt} \def\e{\hbox{\large \it e}}
\def\half{\frac{1}{2}} \def\fr{\frac} \def\arr{\rightarrow}

\def\P{{\rm P}} \def\QG{{\rm QG}} \def\pl{{\rm pl}} \def\M{{\rm M}}        
   \def\V3{{\rm V}_3} \def\EH{{\rm EH}}

 \def\hg{{\hat g}} \def\hDelta{{\hat \Delta}} \def\hnb{{\hat \nabla}}  \def\hR{{\hat R}}   \def\hG{{\hat G}} 

\def\bnb{{\bar \nabla}} \def\bg{{\bar g}} \def\bDelta{{\bar \Delta}} \def\bR{{\bar R}}
\def\bE{{\bar E}} \def\bG{{\bar G}} \def\bC{{\bar C}}

   \def\tildet{\tilde{t}}   

\def\lap3{~| \!\!\! \partial^2} \def\dlap3{~| \!\!\! \partial^4}

\begin{document}

\begin{titlepage}

\begin{flushright}
{\sc July 2009}
\end{flushright}

\vspace{1mm}

\begin{center}
{\Large {\bf Renormalizable 4D Quantum Gravity as A Perturbed Theory from CFT}}
\end{center}

\vspace{1mm}

\begin{center}
{\sc Ken-ji Hamada\footnote{E-mail address: hamada@post.kek.jp}}
\end{center}

\begin{center}
{\it Institute of Particle and Nuclear Studies, KEK, Tsukuba 305-0801, Japan} \\ and \\
{\it Department of Particle and Nuclear Physics, The Graduate University for Advanced Studies (Sokendai), Tsukuba 305-0801, Japan}
\end{center}

\begin{abstract}
We study the renormalizable quantum gravity formulated as a perturbed theory from conformal field theory (CFT) on the basis of conformal gravity in four dimensions. The conformal mode in the metric field is managed non-perturbatively without introducing its own coupling constant so that conformal symmetry becomes exact quantum mechanically as a part of diffeomorphism invariance. The traceless tensor mode is handled in the perturbation with a dimensionless coupling constant indicating asymptotic freedom, which measures a degree of deviation from CFT. Higher order renormalization is carried out using dimensional regularization, in which the Wess-Zumino integrability condition is applied to reduce indefiniteness existing in higher-derivative actions. The effective action of quantum gravity improved by renormalization group is obtained. We then make clear that conformal anomalies are indispensable quantities to preserve diffeomorphism invariance. Anomalous scaling dimensions of the cosmological constant and the Planck mass are calculated. The effective cosmological constant is obtained in the large number limit of matter fields. 
\end{abstract}

\end{titlepage}


\newpage


\section{Introduction}
\setcounter{equation}{0}
\noindent

At very high energies beyond the Planck scale, the space-time would be totally fluctuating quantum mechanically so that geometry loses its classical meanings \cite{dewitt,veltmann, weinberg79}. However, if we apply the Einstein gravity for the Planck scale phenomena, we will encounter fatal difficulties, such as the black hole singularity and divergences in the canonical quantization procedure. Historically, there are many works tackled the divergence problem introducing four-derivative terms in the action of gravity \cite{ud,deser,weinberg74,stelle,jt,tomboulis,ft,ft-report}, because the gravitational coupling constant becomes dimensionless, and at the same time we can avoid the unbounded problem of the action. Nevertheless, any attempt in a fully perturbative approach has not been succeeded. So, most researchers in this field feel the necessity of quantizing gravity in a non-perturbative manner.

Conformal field theory (CFT) is a reliable candidate for such a non-perturbative quantum field theory.  Conformal invariance seems to be crucial to formulate quantum theory of gravity free from such difficulties, because in the early stage of the universe beyond the Planck scale, a scale-invariant space-time would be realized as a part of quantum diffeomorphism invariance, so called background metric independence. Cosmologically, it seems to be natural to consider that a scale invariance of the primordial fluctuations originates from the conformal symmetry. The renormalizable quantum theory of gravity we are going to discuss is that formulated as a perturbed theory from such CFT, as in Fig.\ref{perturbation scheme}.

\begin{figure}[h]
\begin{center}
\begin{picture}(300,80)(10,0)
\CArc(80,50)(20,0,360)
\Text(15,50)[]{Our model ~$=$}\Text(80,50)[]{CFT}\Text(230,50)[]{$+$ ~ perturbations by a single coupling constant}
\CArc(80,0)(20,0,360)
\Text(11,0)[]{Early models ~$=$}\Text(80,0)[]{Free}\Text(223,0)[]{$+$ ~ perturbations by two coupling constants}
\end{picture} 
\end{center}
\caption{\label{perturbation scheme}{\small Our model is formulated as a perturbed theory from CFT by a single coupling constant $t$, while the early models in the 1970-80s were formulated as a perturbed theory from free fields of both the conformal mode and the traceless tensor mode by introducing the coupling constant each.}}
\end{figure}

To define the action, we employ four-derivative conformally invariant quantities in addition to the ordinary lower-derivative actions. It is widely believed that the Weyl tensor denoted by $C_{\mu\nu\lam\s}$ plays a significant role in the early universe where it would vanish as required for the inflation. Since the Weyl tensor includes the Riemann-Christoffel curvature tensor, the vanishing of this tensor means that the singular configuration such as a black hole is excluded. So, we consider the perturbation about the vanishing Weyl tensor introducing a dimensionless coupling constant, $t$, that will be justified by the asymptotically free behavior.

Thus the renormalizable quantum gravity is defined by the following dimensionless action \cite{hamada02, hamada08, nova}:
\begin{eqnarray}
   I = \int d^4 x \sq{-g} \biggl\{
         -\fr{1}{t^2} C^2_{\mu\nu\lam\s} -b G_4 
         + \fr{1}{\hbar} \biggl( \fr{1}{16\pi G}R -\Lam +{\cal L}_\M
          \biggr) \biggr\}.
         \label{quantum gravity action}
\end{eqnarray}
The quantization is carried out by the path integral over the metric field with the weight $e^{iI}$. Here, we write the Newton constant as $G$, and the cosmological constant as $\Lam$. The Lagrangian for a matter field action is denoted by ${\cal L}_\M$. The Euler density $G_4$ is another conformally invariant combination defined by
\begin{equation}
     G_4 = R^2_{\mu\nu\lam\s} -4 R^2_{\mu\nu} + R^2.
     \label{G_4}
\end{equation}
The constant $b$ is introduced to renormalize divergences proportional to this term, which is not an independent coupling constant because it does not have a kinetic term.

There is no $R^2$ term in the action, which is commonly introduced as the kinetic term of the conformal mode in the metric field.\footnote{ 
Thus, this model is free from the problem in the early days \cite{ft} that the $R^2$ action with right sign for the positivity does not show the asymptotic freedom.
}  
It is because the type of divergences is restricted by the Wess-Zumino integrability condition \cite{wz} for conformal anomaly \cite{cd,ddi,duff}. Although most of diffeomorphism invariant counterterms pass this condition, the $R^2$ counterterm is forbidden in four dimensions. This condition is expressed as non-renormalization of the conformal mode. In the following sections, we will see  that the condition is indeed preserved at higher orders using dimensional regularization.

The constant $\hbar$ is the Planck constant, which does not appear in front of the four-derivative gravitational actions because, contrary to matter fields, gravitational fields are exactly dimensionless and thus these actions in four dimensions are dimensionless. This implies that the four-derivative gravitational actions describe purely quantum states, and have no classical meanings.

In order to define the perturbation theory in $t$ about a conformally flat configuration satisfying $C_{\mu\nu\lam\s}=0$, the metric field is decomposed into the conformal mode $\phi$, the traceless tensor mode $h^\mu_{~\nu}$, and the background metric $\hg_{\mu\nu}$. The traceless tensor mode is handled in perturbation, while the conformal mode is treated exactly without introducing its own coupling constant as
\begin{equation}
    g_{\mu\nu}= e^{2\phi}\bg_{\mu\nu}
     \label{metric decomposition}
\end{equation}
and
\begin{equation}
    \bg_{\mu\nu}=\bigl( \hg e^{th} \bigr)_{\mu\nu}
     =\hg_{\mu\lam} \left( \dl^\lam_{~\nu} +th^\lam_{~\nu} 
         +\fr{t^2}{2} ( h^2 )^\lam_{~\nu} +\cdots \right),
        \label{traceless metric field}
\end{equation}
where $tr(h)=h^\lam_{~\lam}=0$. The contraction of the indices of $h^\mu_{~\nu}$ is done by using the background metric. In the following, gravitational quantities with hat and bar on them are defined in terms of the metric $\hg_{\mu\nu}$ and $\bg_{\mu\nu}$, respectively.

Diffeomorphism invariance is defined by the transformation $\dl_\xi g_{\mu\nu}=g_{\mu\lam}\nb_\nu \xi^\lam + g_{\nu\lam}\nb_\mu \xi^\lam$. Under the decompositions of (\ref{metric decomposition}) and (\ref{traceless metric field}), each mode of the metric field transforms independently as
\begin{eqnarray}
   \dl_\xi h_{\mu\nu} &=& \fr{1}{t} \left( \hnb_\mu \xi_\nu +\hnb_\nu \xi_\mu
           - \fr{1}{2} \hg_{\mu\nu} \hnb_\lam \xi^\lam \right)
           + \xi^\lam \hnb_\lam h_{\mu\nu}   
             \nonumber \\
     &&      + \fr{1}{2} h_{\mu\lam} \left( \hnb_\nu \xi^\lam 
                  - \hnb^\lam \xi_\nu \right) 
             + \fr{1}{2} h_{\nu\lam} \left( \hnb_\mu \xi^\lam 
                  - \hnb^\lam \xi_\mu \right) 
           + o(t h^2),
            \nonumber \\
   \dl_\xi \phi &=& \xi^\lam \hnb_\lam \phi  
                       + \fr{1}{4} \hnb_\lam \xi^\lam ,     
               \label{diffeomorphism for traceless mode}
\end{eqnarray}
where the covariant vector $\xi_\mu$ is defined using the background metric as $\xi_\mu =\hg_{\mu\nu} \xi^\nu$.

Consider the vanishing limit of the coupling constant to discuss properties of physical states. At the limit, there are two types of diffeomorphism symmetry: gauge invariance for the kinetic term of the Weyl action and conformal invariance we emphasize here \cite{hamada08}. The former is described by introducing the gauge parameter $\kappa^\mu=\xi^\mu/t$ and taking the limit $t \to 0$ with leaving $\kappa^\mu$ finite. From the transformation (\ref{diffeomorphism for traceless mode}), the diffeomorphism is then expressed as $\dl_\kappa h_{\mu\nu} = \hnb_\mu \kappa_\nu + \hnb_\nu \kappa_\mu - \hg_{\mu\nu} \hnb_\lam \kappa^\lam/2$, while other fields do not transform under the limit such as $\dl_\kappa \phi =0$, because their transformations become of order of $t$ in the expansion using $\kappa^\mu$.

The conformal invariance is now given by the diffeomorphism symmetry with a gauge parameter $\xi^\mu =\zeta^\mu$ satisfying the conformal Killing equation
\begin{equation}
    \hnb_\mu \zeta_\nu + \hnb_\nu \zeta_\mu 
                 - \fr{1}{2} \hg_{\mu\nu} \hnb_\lam \zeta^\lam =0.
           \label{conformal Killing equation}
\end{equation}
Since the lowest term of the transformation of $h_{\mu\nu}$ (\ref{diffeomorphism for traceless mode}) vanishes in this case, the second term becomes effective such that the kinetic term of the Weyl action becomes invariant under the conformal transformation
\begin{equation}
    \dl_\zeta h_{\mu\nu} = \zeta^\lam \hnb_\lam h_{\mu\nu} 
              + \half h_{\mu\lam} \left( \hnb_\nu \zeta^\lam - \hnb^\lam \zeta_\nu \right)
              + \half h_{\nu\lam} \left( \hnb_\mu \zeta^\lam - \hnb^\lam \zeta_\mu \right)  
       \label{traceless mode conformal transformation}
\end{equation}
without taking into account self-interaction terms. The transformation laws of other fields are also obtained in this way such as $\dl_\zeta \phi = \zeta^\lam \hnb_\lam \phi + \hnb_\lam \zeta^\lam/4$.

The dynamics of the traceless tensor mode is governed by the Weyl action, while that of the conformal mode is induced from the path integral measure as in the case of two dimensional quantum gravity \cite{polyakov,kpz,dk,david,seiberg}. We change the path integral measures from the diffeomorphism invariant measures to the practical measures defined on the background. In order to preserve diffeomorphism invariance, the Wess-Zumino action $S$ related to conformal anomalies is necessary as the Jacobian, and the partition function is expressed as 
\begin{equation}
     Z = \int \fr{[d\phi dh \cdots]_\hg}{\rm Vol(diff.)} 
          \exp \left\{ iS(\phi,\bg)+ iI \right\}.
\end{equation}
The induced action $S$ contains the kinetic term of the conformal mode. At the vanishing limit of the coupling constant, $S$ is given by the Riegert action \cite{riegert}
\begin{equation}
     - \fr{b_1}{(4\pi)^2} \int d^4 x \sq{-\hg} \left\{
        2 \phi \hDelta_4 \phi +  \left( \hG_4 - \fr{2}{3} \hnb^2 \hR \right) \phi \right\}, 
        \label{riegert action in introduction}
\end{equation}
where $\sq{-g}\Delta_4$ is a confromally invariant fourth-order operator defined later. The coefficient $b_1$ is a constant with correct sign of $b_1 > 0$.

The Riegert action is a four-dimensional counter quantity of the so-called Liouville-Polyakov action in two dimensions \cite{polyakov}, related to the conformal anomaly proportional to the Euler density. This conformal anomaly, against its name, plays an essential role to make the conformal symmetry exact quantum mechanically at the vanishing coupling limit \cite{am,amm92,amm97,hs,hh,hamada05,hamada08}. By solving the conformal invariance condition, as a generalization of the Virasoro condition in two dimensions, we can obtain diffeomorphism invariant physical states \cite{hh,hamada05,hamada08}.\footnote{ 
It has been shown that the conformal transformations of $h_{\mu\nu}$ and $\phi$ are generated by the Weyl and Riegert (not $R^2$) actions, respectively, and then the negative-metric modes are indeed necessary to form the closed algebra of conformal symmetry quantum mechanically \cite{hamada08}.} 

There also exist ordinary coupling-dependent conformal anomalies following the dynamical mass scale that breaks conformal invariance \cite{cd,ddi,duff}, which play a significant role when we consider the transition from quantum space-time to our real world. In any case, all conformal anomalies arise to preserve diffeomorphism invariance.


\section{Dimensional Regularization}
\setcounter{equation}{0}
\noindent

Among various regularization schemes to extract UV divergences, we use dimensional regularization \cite{tv,veltmann} because it is a systematic and manifestly diffeomorphism invariant method to compute higher order corrections. In this section, we give the definition of renormalizable quantum gravity in this regularization scheme \cite{hamada02}.

In dimensional regularization, the results are independent of how the path integral measure is chosen. It is an advantage of this regularization scheme. Namely, in exactly four dimensional scheme such as the DeWitt-Schwinger method has to evaluate the divergent quantity $\dl^{(4)}(0)=\langle x^\pp | x \rangle |_{x^\pp \arr x}$ that is a contribution from the measure, while in dimensional regularization such a quantity vanishes exactly as $\dl^{(D)}(0)=\int d^D k = 0$. The measure contribution is, instead, included between $D$ and $4$ dimensions, which remains as a finite quantity when the four-dimensional limit is taken after the UV divergence is removed. Therefore, renormalization has to be carried out with care to the conformal-mode dependence in $D$ dimensions.

\subsection{$D$-dimensional actions}
\noindent

First of all, we summarize the $D$-dimensional quantities used in dimensional regularization. How to settle the problem of indefiniteness arising when we generalize the action (\ref{quantum gravity action}) to arbitrary dimensions will be discussed in the following subsection.

We consider quantum theory of gravity coupled to QED with $n_F$ fermions as was discussed in \cite{hamada02}, because QED is the simplest prototype of gauge field theories we meet in practice. The renormalizable $D$-dimensional action near four dimensions is given by
\begin{equation}
    I=\int d^D x \sq{g} \biggl\{
        \fr{1}{t^2} C_{\mu\nu\lam\s}^2 + b G_D +\fr{1}{4}F^2_{\mu\nu}
        + \sum_{j=1}^{n_F} i{\bar \psi}_jD\!\!\!\!/ \psi_j
        -\fr{M_\P^2}{2} R + \Lam  \biggr\} ,
       \label{D dim. action}
\end{equation}
where $M_\P=1/\sq{8\pi G}$ is the reduced Planck mass and $\hbar$ is taken to be unity. We work in the Euclidean space obtained by the Wick rotation on the flat background to evaluate the Feynman integrals.

The first term is the square of the $D$-dimensional Weyl tensor given by
\begin{equation}
    C_{\mu\nu\lam\s}^2 = R_{\mu\nu\lam\s}R^{\mu\nu\lam\s}
           -\fr{4}{D-2}R_{\mu\nu}R^{\mu\nu}
           +\fr{2}{(D-1)(D-2)}R^2 .
        \label{D dim. Weyl action}
\end{equation}
The second term is the $D$-dimensional generalization of the Euler density defined by
\begin{equation}
    G_D =G_4 +\fr{(D-3)^2(D-4)}{(D-1)^2(D-2)}R^2,
               \label{G_D}
\end{equation}
where $G_4$ is the combination (\ref{G_4}) used in four dimensions.

The Dirac operator is defined by $D\!\!\!\!/=e^{\mu\a}\gm_{\a}D_{\mu}$, where $e_{\mu}^{~\a}$ is the vierbein field in $D$ dimensions satisfying the relations $e_{\mu}^{~\a}e_{\nu\a}=g_{\mu\nu}$ and $e_{\mu\a}e^{\mu}_{~\b}=\dl_{\a\b}$, and the Dirac's gamma matrix is normalized to be $\{ \gm_{\a}, \gm_{\b} \}=-2\dl_{\a\b}$. The covariant derivative for fermions is defined by $D_{\mu}=\pd_{\mu}+\half \om_{\mu\a\b}\Sigma^{\a\b}+ie A_{\mu}$, where the connection 1-form and the Lorentz generator are defined by $\om_{\mu\a\b}=e^{\nu}_{~\a}(\pd_{\mu}e_{\nu\b}-\Gamma^{\lam}_{~\mu\nu}e_{\lam\b})$ and $\Sigma^{\a\b}=-\fr{1}{4}[\gm^{\a},\gm^{\b}]$, respectively.

Renormalization can be carried out by replacing the bare quantities in the action $I$ with the renormalized quantities multiplied by the renormalization factors. To make the model finite, we need the following renormalization factors:
\begin{equation}
    A_\mu = Z_3^{1/2}A^r_\mu, \quad
    \psi_j=Z_2^{1/2}\psi^r_j, \quad
    h_{\mu\nu}=Z_h^{1/2}h^r_{\mu\nu} 
    \label{renormalization factors for fields}
\end{equation}
for field variables and 
\begin{equation}
    e=Z_e e_r, \quad t=Z_t t_r 
    \label{renormalization factors for couplings}
\end{equation}
for the coupling constants of QED and the traceless tensor field. The Ward-Takahashi identity holds even if QED couples with quantized gravity so that $Z_e=Z_3^{-1/2}$ is satisfied.

The most remarkable property of this model is that the renormalization factor for the conformal mode is unity:
\begin{equation}
           Z_\phi = 1 .
             \label{Z_phi}
\end{equation}
This comes from the fact that we treat the conformal mode exactly without introducing its own coupling constant. Hence, the verification of this equation can be used as a consistency check of renormalizablity at higher orders.

In dimensional regularization, the UV divergences arise as negative powers of $D-4$, and thus the renormalization factor is expanded as 
\begin{equation}
    Z_3 = 1 + \fr{x_1}{D-4} + \fr{x_2}{(D-4)^2} +\cdots,
     \label{Z_3}
\end{equation}
where the residues $x_n$ are given by functions of the renormalized coupling constants, $e_r$ and $t_r$.

The UV divergences related to the $G_D$ term are renormalized by writing $b$ as
\begin{equation}
      b= \fr{1}{(4\pi)^2}\sum^{\infty}_{n=1}\fr{b_n}{(D-4)^n}, 
              \label{b series}
\end{equation}
because the lowest (tree) part, $b_0 \sq{g}G_D$, does not contain the kinetic term for gravitational fields, so that it is not relevant dynamically. Thus, $b$ is not an independent coupling constant. The constant $b_0$ is taken to be vanishing and the residues of $b_n~(n \geq 1)$ are given by functions of the renormalized couplings.

\subsection{The Wess-Zumino integrability condition}
\noindent

In order to obtain the $D$-dimensional gravitational action defined above, we applied the Wess-Zumino integrability condition \cite{wz} for conformal anomalies \cite{bcr,riegert}, by generalizing the condition to $D$ dimensions, as discussed below.

Consider a generic local-form of conformal anomaly given by the Weyl transformation of the effective action as
\begin{equation}
    \dl_\om \Gm = \int d^D x \sq{g} ~\om \Bigl\{ 
        \eta_1 R^2_{\mu\nu\lam\s} +\eta_2 R^2_{\mu\nu} + \eta_3 R^2 
        +\eta_4 \nb^2 \! R + m_1 R + m_2 \Bigr\} ,
              \label{variation of effective action}
\end{equation}
where the Weyl transformation is defined by $\dl_\om g_{\mu\nu}=2\om g_{\mu\nu}$. Since conformal anomalies arise together with UV divergences, it is a possible candidate for the counterterm to renormalize divergences, or the bare action. The integrability condition is now defined such that two independent Weyl transformations commute as
\begin{eqnarray}
   [\dl_{\om_1}, \dl_{\om_2}] \Gm  &=& 2 \left[ 4\eta_1 +D \eta_2 +4(D-1)\eta_3 + (D-4)\eta_4 \right] 
                        \nonumber \\
             && \times \int d^D x \sq{g} R \om_{[1} \nb^2 \om_{2]} = 0,
                         \label{D dim. integrability condition}
\end{eqnarray}
where the anti-symmetric product is denoted as $a_{[\mu}b_{\nu]}=(a_\mu b_\nu -a_\nu b_\mu)/2$. Thus, the integrability gives a constraint on the form of the fourth-order action, while the Einstein action and the cosmological constant term are trivially integrable. This condition indicates the renormalizability of quantum gravity such that the effective action exists.

One of the combination satisfying the integrability condition (\ref{D dim. integrability condition}) is the $D$-dimensional Weyl action (\ref{D dim. Weyl action}). The other combinations are $G_4$ and 
\begin{equation}
     M_D = \nb^2 R - \fr{D-4}{4(D-1)}R^2.
\end{equation}
$M_D$ is an integrable generalization of the trivial conformal anomaly $\nb^2 R$, which is no longer trivial in $D$ dimensions. Here, note that in exactly four dimensions the integrable quantities are just two of the square of the Weyl tensor and the Euler density, apart from the trivial term with the parameter $\eta_4$, as given by (\ref{quantum gravity action}) in Introduction.

We further reduce the ambiguity in the $D$-dimensional action by requiring that it satisfies a property analogous to that the action of two-dimensional quantum gravity near two dimensions has. The action of two-dimensional quantum gravity is given by the scalar curvature. It is a unique second-order action that is integrable even in $D$ dimensions. The expansion of the action about $2$ dimensions is given by 
\begin{equation}
     \int d^D x \hbox{$\sq{g}$} R = \sum_{n=0}^\infty \fr{(D-2)^n}{n!}S_n^{(2)}(\phi,\bg) .
\end{equation}
Each term $S_n^{(2)}$ has the following form:
\begin{equation}
     S_n^{(2)}(\phi,\bg) = \int d^D x \hbox{$\sq{\bg}$} 
          \left\{ \phi^n \bDelta_2 \phi + \bR \phi^n + o(\phi^n) \right\},
\end{equation}
where $\sq{g}\Delta_2 = \sq{g}(-\nb^2)$ is the second-order differential operator that becomes conformally invariant at $D=2$, and $o(\phi^n)$ denotes the terms given by the at most $n$-th product of the $\phi$ field. The action $S_1^{(2)}$ is the well-known Liouville-Polyakov action in two dimensional quantum gravity \cite{polyakov,kpz,dk,david,seiberg}.

We impose the condition that the similar property is satisfied for the four-dimensional quantum gravity near four dimensions. Consider an integrable combination $E_D=G_4+\eta M_D$ and the expansion of it about four dimensions. We look for the parameter $\eta$ such that the volume integral of $E_D$ has the following expansion series:
\begin{equation}
      \int d^D x \sq{g}E_D  = \sum^{\infty}_{n=0} \fr{(D-4)^n}{n!}S_n(\phi,\bg),
             \label{expansion of E_D}
\end{equation}
where each term $S_n$ has the property
\begin{equation}
     S_n (\phi,\bg)
     = \int d^D x \sq{\bg} \left\{ 2 \phi^n \bDelta_4 \phi + \bE_4 \phi^n
        + o(\phi^n) \right\}.
\end{equation}
The fourth order differential operator $\sq{g}\Delta_4$ acting on a scalar variable is defined by
\begin{equation}
     \Delta_4 =  \nb^4 + 2R^{\mu\nu}\nb_\mu \nb_\nu - \fr{2}{3}R \nb^2 
                + \fr{1}{3} \nb^\mu R \nb_\mu,
\end{equation}
which becomes conformally invariant at four dimensions and the quantity $E_4$ represents the modified Euler density in four dimensions given by
\begin{equation}
    E_4 = G_4 - \fr{2}{3} \nb^2 R .
\end{equation}
This combination is defined so as to satisfy the relation $\sq{g}E_4=\sq{\bg}(4\bDelta_4 \phi +\bE_4)$ in four dimensions, similar to the relation $\sq{g}R=\sq{\bg}(2\bDelta_2 \phi +\bR)$ for the Euler density in two dimensions. Thus, the similarity between two and four dimensional models is now apparent. The lowest action $S_1$ is nothing but the Riegert action given by (\ref{riegert action in introduction}).

This condition determines the parameter $\eta$ uniquely to be $-4(D-3)^2/(D-1)(D-2)$. This value reduces to $-2/3$ in four dimensions as the modified Euler density $E_4$ holds. Thus, we obtain the combination $E_D$ as the $D$-dimensional generalization of the modified Euler density,
\begin{equation}
     E_D = G_D -\fr{4(D-3)^2}{(D-1)(D-2)} \nabla^2 R ,
          \label{E_D}
\end{equation}
where $G_D$ is the quantity we seek, given by (\ref{G_D}).

The action $G_D$ (\ref{G_D}) reproduces the Hathrell's result on conformal anomalies computed at the three-loop level of order $e^6_r$ in the curved space \cite{hathrell-QED}. In order to renormalize the UV divergence, he use the following counterterm
\begin{equation}
      b G_4 + c H^2
\end{equation}
in addition to the $D$-dimensional Weyl action (\ref{D dim. Weyl action}), where $H=R/(D-1)$. The gravitational action $G_D$ we found means that the two coefficients should satisfy the relationship 
\begin{equation}
       c=\fr{(D-3)^2 (D-4)}{(D-2)}b.
\end{equation}
Expanding $c$ in the Laurent series of $D-4$ as in the case of $b$ (\ref{b series}), it represents the relationship among the residues,
\begin{equation}
      c_1=\fr{(D-3)^2}{D-2}b_2 = \half b_2 + o(D-4).
      \label{Hathrell relation}
\end{equation}
Hathrell showed that this relationship is indeed satisfied at the order of $e_r^6$ for QED in a curved space-time. He also showed that the same relation holds in the case of conformally coupled scalar field theory with four-point interaction \cite{hathrell-scalar}. Furthermore, it was confirmed in the case of non-abelian gauge theory by Freeman \cite{freeman}. These results indicate that the combination $G_D$ is universal independent of matter contents.

Historically, the Hathrell's results were, at first, understood as an evidence that the Wess-Zumino integrability condition is broken at three-loop level due to the appearance of the residue $c$. The problem is now resolved and it is shown that the universality of his results does support that there is no independent $R^2$ counterterm. This condition is expressed by non-renormalization of the conformal mode (\ref{Z_phi}), and it is indeed confirmed through the whole computations including gravitational loop corrections.


\section{Conformal Anomalies and Effective Actions}
\setcounter{equation}{0}
\noindent

Renormalization is carried out by expanding bare quantities in terms of renormalized quantities, and regarding terms with non-negative powers of $D-4$ as kinetic terms or vertices, and those with negative powers of $D-4$ as counterterms. Here, before carrying out definite calculations of renormalization factors, we discuss how conformal anomalies arise in dimensional regularization and show that they are indeed necessary to preserve diffeomorphism invariance.

We here divide conformal anomalies into two groups: coupling-dependent and coupling-independent ones. The former is an ordinary conformal anomaly that arises following beta functions for the dynamics of gauge field and traceless tensor field, which represents a violation of the conformal symmetry by a dynamical scale. The latter appears as a coupling-independent part of the conformal anomaly proportional to the Euler term. Contrary to the former, it plays an important role to make conformal symmetry exact as a part of the diffeomorphism invariance, as mentioned in Introduction.

To begin with, we consider the QED sector, which produces conformal anomalies in the former group. Using the Laurent expansion of $Z_3$ (\ref{Z_3}), the gauge field action is expanded as
\begin{eqnarray}
  &&  \fr{1}{4} \int d^D x \sq{g}F^2_{\mu\nu}
             \nonumber  \\
  &&
     =\fr{1}{4}Z_3\int d^D x e^{(D-4)\phi} 
         {\bar F}^r_{\mu\nu} {\bar F}^r_{\lam\s} \bg^{\mu\lam}\bg^{\nu\s}
              \nonumber   \\
  &&  =\fr{1}{4} \int d^D x \biggl\{
          \biggl(1+ \fr{x_1}{D-4}+\fr{x_2}{(D-4)^2}+\cdots \biggr)
            {\bar F}^r_{\mu\nu} {\bar F}^r_{\lam\s} \bg^{\mu\lam}\bg^{\nu\s}
                \nonumber  \\
  && \qquad\qquad
          +\biggl(D-4 + x_1+\fr{x_2}{D-4}+\cdots \biggr)
            \phi  {\bar F}^r_{\mu\nu} {\bar F}^r_{\lam\s} \bg^{\mu\lam}\bg^{\nu\s}
                \nonumber  \\
  && \qquad\qquad
          +\fr{1}{2}\biggl( (D-4)^2 + (D-4)x_1+ x_2+\cdots \biggr)
              \phi^2 {\bar F}^r_{\mu\nu} {\bar F}^r_{\lam\s} \bg^{\mu\lam}\bg^{\nu\s} 
                \nonumber \\
   && \qquad\qquad
             + \cdots \biggr\}, 
          \label{D dim. gauge action}
\end{eqnarray}
where ${\bar F}^r_{\mu\nu} = \bnb_\mu A_\nu^r-\bnb_\nu A_\mu^r ~(= \pd_\mu A_\nu^r-\pd_\nu A_\mu^r)$. Expanding the bare metric field $\bg_{\mu\nu}$ in terms of the renormalized coupling constant $t_r$ and the renormalized traceless tensor field $h^r_{\mu\nu}$, we obtain gravitational vertices and their counterterms. Here, note that as mentioned before the conformal mode $\phi$ is not renormalized so that $Z_\phi=1$.

The first line in the Laurent expansion (\ref{D dim. gauge action}) yields the kinetic term and counterterms of the gauge field as usual. Thus, the residues $x_n$ are determined from the UV divergences proportional to the kinetic term.

The new vertices and counterterms of the form $\phi {\bar F}^{r 2}_{\mu\nu}$ in the second line, which represents the Wess-Zumino action for the conformal anomaly, are induced by the residue $x_1$ as a quantum effect. In general, the residue $x_n$ induces the new vertices and counterterms of the form $\phi^n {\bar F}^{r 2}_{\mu\nu}$ as in the third line and below. Within the QED calculation, the residue $x_1$ is given by
\begin{equation}
   x_1 = \fr{8n_F}{3}\fr{e_r^2}{(4\pi)^2} + 4n_F \fr{e_r^4}{(4\pi)^4}.
        \label{residue of QED}
\end{equation}

At this order, there is a logarithmic non-local term of $\log (k^2/\mu^2)$ as a finite correction, where $k^2~(=k_\mu k_\nu \dl^{\mu\nu})$ is the square of momentum in the flat background and $\mu$ is an arbitrary mass scale. The coefficient of this term is proportional to 
\begin{equation}
      y_1 = \fr{8n_F}{3}\fr{e_r^2}{(4\pi)^2} + 8 n_F \fr{e_r^4}{(4\pi)^4},
\end{equation}
corresponding to the beta function $\b_e/e_r = y_1/2$. Here, the coefficient of the one-loop correction is the same to that in $x_1$, while the two-loop correction is the twice of that.

\begin{figure}[h]
\begin{center}
\begin{picture}(500,60)(120,30)
\Photon(200,50)(220,50){2}{3}
\DashArrowArc(250,50)(30,0,180){1}
     \DashArrowArc(250,50)(30,180,360){1}
  \PhotonArc(250,80)(30,210,330){2}{8}     
    \Line(250,50)(250,65) 
    \Vertex(250,50){1}\Text(250,40)[]{$\eps$}
\Photon(280,50)(300,50){2}{3} 

\Photon(330,50)(350,50){2}{3}
\DashArrowArc(380,50)(30,30,210){1}
     \DashArrowArc(380,50)(30,210,390){1}
  \Photon(380,80)(380,20){2}{7}     
    \Line(380,50)(395,50) 
    \Vertex(380,50){1}\Text(378,50)[r]{$\eps$}
\Photon(410,50)(430,50){2}{3} 
\end{picture} 
\end{center}
\caption{\label{finite QED loop}{\small Finite corrections to the vertex $\phi {\bar F}^{r 2}_{\mu\nu}$ at order $e_r^4$. The wavy line and the dashed line with arrows denote the photon and the fermion propagators, respectively. The solid line denotes the conformal mode and $\eps$ at the vertices is $(4-D)/2$.}}
\end{figure}
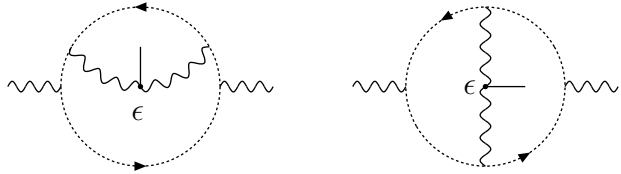

After carrying out renormalization and taking the four dimensional limit, we obtain the following effective action in the momentum space:
\begin{equation}
       \Gm_{\rm QED} =  \Biggl\{ 
         1 -\fr{y_1}{2} \log \left( \fr{k^2}{\mu^2} \right) 
          + x_1 \phi + 4n_F \fr{e_r^4}{(4\pi)^4}\phi \Biggr\} 
            \fr{1}{4} {\bar F}^{r 2}_{\mu\nu}, 
\end{equation}
where, for simplicity, only the zero-momentum mode of $\phi$ is considered. The first and second terms in the right-hand side are the tree part and finite loop corrections, respectively. The third term is the induced vertex by the residue $x_1$, as mentioned above, while the fourth term comes from the sum of loop corrections depicted in Fig.\ref{finite QED loop}, which gives the finite contribution because the two-loop photon self-energy of QED yields at most a simple pole, which cancels out $D-4$ on the vertex coming from the first term in the second line of the Laurent expansion (\ref{D dim. gauge action}).

Introducing the physical momentum defined on the full metric by
\begin{equation}
    p^2 = k^2/e^{2\phi},
    \label{physical momentum} 
\end{equation}
the effective action can be written as
\begin{equation}
      \Gm_{\rm QED} =  \left\{   1 -\fr{y_1}{2} \log \left( \fr{p^2}{\mu^2} \right) 
                              \right\} \fr{1}{4}  \sq{g_r}F^{r 2}_{\mu\nu}.
\end{equation}
In this way, we can demonstrate that the effective action is written in terms of the full metric field $g^r_{\mu\nu}$.

At higher orders, following the non-local term $\log^n (k^2/\mu^2)$, the Wess-Zumino action like $\sq{g_r}\phi^n F_{\mu\nu}^{r 2}$ is induced to make the effective action diffeomorphism invariant.

Similarly, the Weyl action can be expanded in terms of the renormalized quantities using expression
\begin{equation}
     \fr{1}{t^2} \int d^D x \hbox{$\sq{g}$} C_{\mu\nu\lam\s}^2
     = \fr{1}{t^2} \int d^D x \hbox{$\sq{\bg}$} e^{(D-4)\phi}{\bar C}_{\mu\nu\lam\s}^2 
\end{equation}
and the renormalization factors of $Z_t$ and $Z_h$. In the same manner as mentioned above, new vertices and counterterms of the type $\phi^n \bC_{\mu\nu\lam\s}^2$ are induced. Although the expression becomes more complicated, the expansion is straightforward.

The non-local term $\b_0 \log(k^2/\mu^2)$ with $\b_0 >0$ arises in connection with the UV divergence of simple pole determining the beta function to be $\b_t=-\b_0t_r^3$. The divergence also induces the Wess-Zumino action for the conformal anomaly of the type $\phi {\bar C}^{r 2}_{\mu\nu\lam\s}$, and thus the effective action is written as 
\begin{eqnarray}
      \Gm_{\rm W} &=& \left\{ 
        \fr{1}{t_r^2} -2\b_0 \phi + \b_0 \log \left( \fr{k^2}{\mu^2} \right) 
                \right\} {\bar C}_{\mu\nu\lam\s}^{r 2}
            \nonumber \\
    &=& \fr{1}{\overline{t}_r^2(p)} \sq{g_r} C_{\mu\nu\lam\s}^{r 2} .
\end{eqnarray}
The function $\overline{t}_r^2(p)$ is the running coupling constant obtained by collecting the terms in the braces as
\begin{equation}
    \overline{t}_r^2(p)= \fr{1}{\b_0 \log (p^2/\Lam_\QG^2)}.
    \label{running coupling constant}
\end{equation}
Here, $p$ is a physical momentum defined by (\ref{physical momentum}) and the parameter $\Lam_\QG = \mu \exp\{ - 1/2\b_0t_r^2\}$ is the new dynamical energy scale. The Wess-Zumino action like $\sq{g_r}\phi^n C_{\mu\nu\lam\s}^{r 2}$ also corresponds to the non-local term $\log^n (k^2/\mu^2)$ at higher orders. The renormalization group equation will be discussed in Section 6.

Next, we consider the Euler term. Using the expansion of $b$ (\ref{b series}), we obtain the following Laurent expansion: 
\begin{eqnarray}
    && b\int d^D x \sq{g} G_D
          \nonumber   \\
    && = \fr{1}{(4\pi)^2} \int d^D x \biggl\{
         \biggl( \fr{b_1}{D-4}+\fr{b_2}{(D-4)^2}
                 + \cdots \biggr) \bG_4
            \nonumber  \\
    && \qquad\qquad
         + \biggl( b_1 +\fr{b_2}{D-4} +\cdots \biggr)
          \biggl( 2 \phi \bDelta_4 \phi +\bE_4 \phi
                      +\fr{1}{18}\bR^2 \biggr)
             \nonumber   \\
    && \qquad
        +\half \Bigl( (D-4)b_1 +b_2 +\cdots \Bigr)
           \Bigl( 2\phi^2 \bDelta_4 \phi +\bE_4 \phi^2
                             +\cdots  \Bigr)
         +\cdots \biggr\}.
         \label{Laurent expansion of bG_D}
\end{eqnarray}
The bare metric field is further expanded in terms of the renormalized variables using the renormalization factors. The residues $b_n$ are determined by the UV divergences proportional to $\bG_4$. The residue $b_1$ produces the kinetic term of the conformal mode, or the Riegert action $S_1(\phi,\bg_r)= \int d^4 x \{2 \phi \bDelta^r_4 \phi +\bE^r_4 \phi + \bR_r^2/18\}$. The third line in the expansion represents new types of the Wess-Zumino action induced at higher orders.

Although the residues $b_n$ are, in general, the functions of the coupling constants, the lowest $b_1$ includes the constant term independent of the couplings. So, we separate it into the constant term and the coupling-dependent term as
\begin{equation}
    b_1(t_r,e_r) = b_1 + b_1^\pp(t_r,e_r).
\end{equation}
In the following, $b_1$ represents the constant term. For $n \geq 2$, there is no such a constant term. The $b_1$ term is necessary to realize the conformal invariance at the vanishing limit of the coupling constant.

We first discuss the lowest part of the effective action with the coefficient $b_1$. The non-local part of the effective action arises as a finite loop correction following the UV divergence proportional to $b_1 \bG^r_4$. So, we seek the effective action that yields $b_1 \bG^r_4$ under the conformal variation of the metric field $\bg^r_{\mu\nu}$, which is given by
\begin{equation}
   W_G (\bg_r) =\fr{b_1}{(4\pi)^2} \int d^4 x \left\{ \fr{1}{8} 
    \bE_4^r \fr{1}{\bDelta^r_4} \bE_4^r -\fr{1}{18}\bR_r^2 \right\} .
\end{equation}
The first term in right-hand side is the so-called non-local Riegert action defined on the metric field $\bg^r_{\mu\nu}$. Since $\bG^r_4$ has no two-point function, this non-local action is determined through the three-point function of the traceless tensor field. The $\bR_r^2$ term guarantees that $W_G(\bg_r)$ has no two-point function of the traceless tensor field when expanding about the flat background.

The diffeomorphism invariant effective action is given by the combination of the induced Wess-Zumino action $b_1S_1$ and the non-local loop correction $W_G$ as
\begin{equation}
    \fr{b_1}{(4\pi)^2} S_1(\phi,\bg_r) + W_G(\bg_r) 
    = \fr{b_1}{8(4\pi)^2}  \int d^4 x \sq{g_r}  
          E_4^r \fr{1}{\Delta^r_4} E_4^r .
          \label{non-local Riegert action}
\end{equation}
Thus, the $\bR_r^2$ terms cancel out and we obtain the non-local Riegert action  written in terms of the renormalized full metric, which is a four-dimensional counter action of the scale-invariant non-local action of Polyakov in two dimensional quantum gravity.

The conformal anomalies with the coupling-dependent coefficients, $b_1^\pp$ and $b_n ~(n\geq 2)$, lead to a violation of conformal symmetry by the dynamical mass scale as discussed in the QED and Weyl sectors. We here consider higher order corrections to the coefficient $b_1$ in front of the kinetic term $L_{S_1}= 2 \phi \bDelta^r_4 \phi$ by taking the coupling-dependent coefficient of $b_1(t_r)=b_1 ( 1-a_1 t_r^2 + \cdots )$. Replacing the coupling constant $t_r$ with the running coupling constant (\ref{running coupling constant}), we obtain the following effective action in the momentum space: 
\begin{eqnarray}
     \Gm_{\rm R} &=&  \fr{b_1}{(4\pi)^2}
            \left( 1-a_1 \overline{t}_r^2(p) + \cdots \right) L_{S_1} (\phi,\bg_r)
             \nonumber \\
     &=&  \fr{b_1}{(4\pi)^2} \left\{  1- a_1 \left[ t_r^2 +2\b_0 t_r^4 \phi 
          - \b_0 t_r^4 \log \left( \fr{k^2}{\mu^2} \right) +\cdots \right] + \cdots
               \right\} L_{S_1} (\phi,\bg_r) .
            \nonumber \\ 
            \label{Riegert action at higher order}
\end{eqnarray}
This expression indicates that the $\phi^2 \bDelta^r_4 \phi$ term and corresponding non-local term with logarithm arise at the order of $t_r^4$, and therefore the coefficient $b_2$ arises at this order.  Thus, it is understood that the interaction $\phi^n \bDelta^r_4 \phi~(n \geq 2)$ is also induced to guarantee diffeomorphism invariance such that the effective action can be written in terms of the running coupling constant at higher orders.

The dynamical scale parameter $\Lam_\QG$ in the running coupling constant (\ref{running coupling constant}) represents the energy scale where the correlation length becomes short-range and thus the conformal invariance breaks down turning to the classical Einstein phase. If we set the ordering of two mass scales as $m_\pl\gg \Lam_\QG ~(\simeq 10^{17}{\rm GeV})$, where $m_\pl=1/\sq{G}~(\simeq 10^{19}{\rm GeV})$, we obtain an inflationary scenario with a sufficient number of e-foldings driven by quantum gravity effects \cite{hy,hhy,hhsy,hms}.


\section{Residues of Renormalization Factors}
\setcounter{equation}{0}
\noindent

In the previous section, we have seen that the new vertices and counterterms related to conformal anomalies are induced to preserve diffeomorphism invariance. Taking into account these terms, we can compute residues of renormalization factors and the beta functions including gravitational loop corrections order by order of the coupling constants. We here summarize the procedure of calculations and various results for these quantities \cite{hamada02}. We then demonstrate that the conformal mode is indeed not renormalized such that the renormalization factor of this mode is unity.

\subsection{Gauge-fixing of the Weyl action}
\noindent

We first carry out the gauge-fixing of the traceless tensor field to obtain the propagator.  The kinetic term of the Weyl action is given by 
\begin{eqnarray}
    && \fr{1}{t^2}\int d^D x \hbox{$\sq{g}$} C_{\mu\nu\lam\s}^2 
          \nonumber \\
    && = \int d^D x \left\{
          \fr{D-3}{D-2} \left( h_{\mu\nu} \pd^4 h_{\mu\nu} 
         + 2\chi_{\mu} \pd^2 \chi_{\mu} \right)
         - \fr{D-3}{D-1} \chi_{\mu} \pd_{\mu}\pd_{\nu}\chi_{\nu}  
           \right\} ,
\end{eqnarray}
where $\chi_{\mu}=\pd_{\lam}h_{\lam\mu}$ and d'Alembertian on the flat background is denoted by $\pd^2 =\pd_{\lam}\pd_{\lam}$. Here and below, the same lower space-time indices represent the contraction by the flat Euclidean metric $\dl_{\mu\nu}$.

Following the standard procedure of the BRST gauge-fixing \cite{ko}, we introduce the gauge-fixing term and the ghost action for the traceless tensor field and the $U(1)$ gauge field as well, 
\begin{eqnarray}
    I_{\rm GF + FP} &=& \int d^D x \dl_{\rm B} \left\{ 
       \tilde{c}_\mu N_{\mu\nu} \left( \chi_\nu -\fr{\zeta}{2}B_\nu \right)
       + \tilde{c} \left( \pd_\mu A_\mu - \fr{\a}{2} B \right) \right\},
\end{eqnarray}
where $\tilde{c}_\mu$ and $\tilde{c}$ are the anti-ghosts and $B_\mu$ and $B$ are the subsidiary fields. $\dl_{\rm B}$ denotes the BRST transformation. $N_{\mu\nu}$ is a symmetric second-order differential operator. Here, it is defined as 
\begin{equation}
    N_{\mu\nu}=  \fr{2(D-3)}{D-2} \left( 
                 -2 \pd^2\dl_{\mu\nu} + \fr{D-2}{D-1} \pd_{\mu}\pd_{\nu} \right) .
\end{equation}

The BRST transformations for the traceless tensor field and the gauge field are obtained by replacing the gauge parameter of the diffeomorphism $\xi^\mu/t$ with the ghost $c^\mu$ and that of the $U(1)$ gauge transformation with the ghost $c$ as
\begin{eqnarray}
      \dl_{\rm B} h_{\mu\nu} &=& \pd_\mu c_\nu + \pd_\nu c_\mu 
            - \fr{2}{D} \dl_{\mu\nu} \pd_\lam c_\lam + t c_\lam \pd_\lam h_{\mu\nu}
                 \nonumber \\
         && 
            + \fr{t}{2} h_{\mu\lam} \left( \pd_\nu c_\lam - \pd_\lam c_\nu \right) 
            + \fr{t}{2} h_{\nu\lam} \left( \pd_\mu c_\lam - \pd_\lam c_\mu \right) 
            + \cdots,
                 \nonumber \\
      \dl_{\rm B}A_\mu &=& \pd_\mu c 
            + t \left( c_\lam \pd_\lam A_\mu + A_\lam \pd_\mu c_\lam \right).
\end{eqnarray}
The BRST transformations of ghosts, anti-ghosts and subsidiary fields are given by
\begin{eqnarray}
     \dl_{\rm B} c_\mu &=& t c_\lam \pd_\lam c_\mu,
        \nonumber \\
     \dl_{\rm B} c &=& t c_\lam \pd_\lam c,
        \nonumber \\ 
     \dl_{\rm B} \tilde{c}_\mu &=& B_\mu, \qquad \dl_{\rm B} B_\mu = 0,
        \nonumber \\  
     \dl_{\rm B} \tilde{c} &=& B, \qquad \dl_{\rm B} B = 0.
\end{eqnarray}
And also, the BRST transformation of the conformal mode, which does not appear in the gauge-fixing term, is given by
\begin{equation}
          \dl_{\rm B} \phi = t c_\lam \pd_\lam \phi + \fr{t}{D} \pd_\lam c_\lam.
\end{equation}

Using the BRST transformation, the gauge-fixing term and the ghost action are expressed as  
\begin{eqnarray}
    I_{\rm GF + FP} &=& \int d^D x \biggl\{ B_\mu N_{\mu\nu} \chi_\nu 
               - \fr{\zeta}{2} B_\mu N_{\mu\nu} B_\nu
               - \tilde{c}_\mu N_{\mu\nu} \pd_\lam (\dl_{\rm B}h_{\nu\lam})
           \nonumber \\
        && \qquad\qquad
            + B \pd_\mu A_\mu - \fr{\a}{2} B^2 - \tilde{c}\pd_\mu (\dl_{\rm B}A_\mu)
            \biggr\} .
\end{eqnarray}
Furthermore, integrating out the subsidiary fields, we obtain the following gauge-fixing term:\footnote{ 
After integrating out the $B_\mu$ field, the determinant $\det^{-1/2}(N_{\mu\nu})$ arises. If one use the background field method in a curved space-time, one has to evaluate this determinant \cite{ft}.}
\begin{equation}
      I_{\rm GF} = \int d^D x  \left\{
        \fr{1}{2\zeta}  \chi_{\mu} N_{\mu\nu} \chi_{\nu} 
        + \fr{1}{2\a} (\pd_\mu A_\mu )^2 \right\}.
\end{equation}

The renormalization of the gauge-fixing parameters are defined by $\a=Z_3 \a_r$ and $\zeta = Z_h \zeta_r$ such that the counterterms for these kinetic terms have the gauge-invariant forms. We also introduce the renormalization factors for the ghost fields as usual. Here, the choice of $\a_r=1$ and $\zeta_r=1$ is the so-called Feynman gauge. In the following, we use the Feynman gauge.

In the Feynman gauge, the propagator of the traceless tensor field becomes
\begin{equation}
       \langle h^r_{\mu\nu}(k) h^r_{\lam\s}(-k) \rangle
       = \fr{D-2}{2(D-3)} \frac{1}{k^4} I^H_{\mu\nu, \lam\s}, 
\end{equation}
where the projection operator to the traceless mode is given by
\begin{equation}
   I^H_{\mu\nu, \lam\s} 
      = \half \left( \dl_{\mu\lam}\dl_{\nu\s}+ \dl_{\mu\s}\dl_{\nu\lam} \right)
                         - \fr{1}{D}\dl_{\mu\nu}\dl_{\lam\s} ,
\end{equation}
which satisfies the condition $I_H^2=I_H$. The propagator of the traceless tensor field is described by the spiral line.

\subsection{Propagator and vertices in the Riegert action}
\noindent

The kinetic term of the conformal mode is contained in the Riegert action induced following the UV divergence with the residue $b_1$ as shown in (\ref{Laurent expansion of bG_D}),
\begin{equation}
    \int d^D x \fr{b_1}{(4\pi)^2}  
      \biggl\{ 2 \phi \bDelta_4 \phi +\bE_4 \phi 
                      +\fr{1}{18}\bR^2 \biggr\} .
                      \label{D dim. Riegert action}
\end{equation}
From this action we obtain the propagator of the conformal mode,
\begin{equation}
    \langle \phi(k) \phi(-k) \rangle = \fr{(4\pi)^2}{4b_1} \fr{1}{k^4},
\end{equation}
which is described by the solid line.

Expanding the the Riegert action (\ref{D dim. Riegert action}) in the coupling constant $t$, we obtain the following interactions:
\begin{eqnarray}
    &&  L^2_{\phi h} = \fr{b_1}{(4\pi)^2} \left\{ -\frac{2}{3} t 
           \pd^2 \phi \pd_{\mu}\pd_{\nu} h_{\mu\nu} 
           +\fr{1}{18} t^2 
           \bigl(\pd_{\mu}\pd_{\nu}h_{\mu\nu}\bigr)^2 \right\},
                 \nonumber   \\ 
    && L^3_{\phi\phi h} = \fr{2b_1}{(4\pi)^2} t \biggl\{
          2\pd_{\mu} \phi \pd_{\nu} \pd^2 \phi 
          + \frac{4}{3}\pd_{\mu}\pd_{\lam}\phi \pd_{\nu}\pd_{\lam}\phi 
                  \nonumber \\ 
    && \qquad\qquad\qquad 
        -\frac{2}{3}\pd_{\lam}\phi \pd_{\mu}\pd_{\nu}\pd_{\lam} \phi 
          -2\pd_{\mu}\pd_{\nu}\phi \pd^2 \phi 
             \biggr\} h_{\mu\nu} ,                
                  \nonumber  \\  
    && L^4_{\phi\phi hh} = \fr{2 b_1}{(4\pi)^2} t^2 \Bigl\{ 
         \pd^2\phi\pd_{\mu}\pd_{\nu} \phi h_{\mu\lam} h_{\nu\lam} 
         +\pd_{\mu}\pd_{\nu}\phi \pd_{\lam}\pd_{\s}\phi 
               h_{\mu\nu}h_{\lam\s} 
               \nonumber \\ 
    && \qquad\qquad\qquad\qquad 
         + ~\hbox{terms including $\pd h$} ~\Bigr\} .
         \label{Interactions in Wess-Zumino action}       
\end{eqnarray}
Here, we write only the terms needed in the following computation. The first and second terms in $L^2_{\phi h}$ come from the $(\bnb^2 \bR)\phi$ term in $\bE_4 \phi$ and the last $\bR^2$ term, respectively. Both $L^3_{\phi\phi h}$ and $L^4_{\phi\phi h h}$ are derived from the $\phi\bDelta_4 \phi$ term. The bare quantities, $t$ and $h_{\mu\nu}$, are further expanded in terms of the renormalized ones to obtain the vertices and counterterms.

\subsection{On UV and IR divergences}
\noindent

We here give various comments on the regularization. As will be known in the following computation, the number of loops does not match with the order of $\hbar$ in quantum gravity. It is apparent from the fact that the Riegert action as the kinetic term of the conformal mode is induced by a loop effect. In general, it is because four-derivative gravitational actions are dimensionless quantities of the vanishing order of $\hbar$ as mentioned in Introduction.

The lower-derivative gravitational actions such as the Einstein action and the cosmological constant term cannot be regarded as ordinary mass terms, because the conformal mode is now treated exactly without introducing its own coupling constant and thus these actions become diffeomorphism invariant in the form with exponential factors of the conformal mode even at the vanishing coupling limit, contrary to the induced Wess-Zumino actions expanded in polynomials of the conformal mode order by  order of the coupling constants.

Therefore, in order to treat infrared (IR) divergences, we introduce a small fictitious mass $z$ to the gravitational field like a photon mass, namely we replace the momentum dependence of the propagator $1/k^4$ with $1/(k^2 +z^2)^2$. The IR divergence then arises in the form of either $\log z^2$ or negative power of $z^2$. Since such a mass term is not gauge invariant, the IR divergence finally cancels out.

Here, we summarize the quantities used in the following calculations:
\begin{eqnarray}
     D= 4-2\eps, \qquad t_r = {\tilde t}_r \mu^\eps, \qquad e_r = {\tilde e}_r \mu^\eps, 
     \qquad b = {\tilde b} \mu^{-2\eps}
     \label{scale dependence of coupling constant}
\end{eqnarray}
where $\mu$ is an arbitrary mass scale and ${\tilde t}_r$, ${\tilde e}_r$ and ${\tilde b}$ are dimensionless quantities. In the real space, the conformal mode that appears in the exponential factor has to be dimensionless even in $D$ dimensions, while the traceless tensor field has the dimension $\mu^{-\eps}$.

\subsection{Non-renormalization theorem for the conformal mode}
\noindent

To begin with, we give the results for the residues $b_n$. Evaluating the UV divergences proportional to $\bG_4$, the residues have been computed as
\begin{eqnarray}
    {\tilde b}_1 &=& \fr{11 n_F}{360}+\fr{40}{9},  \qquad
    {\tilde b}_1^\pp = -\fr{n_F^2}{6}\fr{{\tilde e}_r^4}{(4\pi)^4} +o({\tilde t}_r^2) ,
        \nonumber \\
    {\tilde b}_2 &=& \fr{2 n_F^3}{9}\fr{{\tilde e}_r^6}{(4\pi)^6}+ o({\tilde t}_r^4) .
               \label{b_1 and b_2}
\end{eqnarray}
The coupling-independent part of the residue of simple pole, $b_1$, is given by the sum of one-loop contributions from three sectors of QED \cite{duff}, the conformal mode \cite{amm92} and the traceless tensor mode \cite{ft}.\footnote{ 
In general, ${\tilde b}_1 = (N_X + 11N_D + 62 N_A)/360+769/180$, where $N_X$, $N_D$ and $N_A$ are the numbers of conformally-coupled  scalar fields, Dirac fermions and gauge fields.
} 
The coupling-dependent residues of simple and double poles, $b_1^\pp$ and $b_2$, are two-loop and three-loop contributions from the QED sector, respectively, computed by Hathrell \cite{hathrell-QED}.

\begin{figure}[h]
\begin{center}
\begin{picture}(430,70)(40,30)
\Line(100,50)(200,50)
\GlueArc(150,50)(30,0,180){3}{10.5}
\Vertex(120,50){1}\Text(120,47)[t]{$b_1 t_r$}
\Vertex(180,50){1}\Text(180,47)[t]{$b_1 t_r$}
\Text(150,20)[]{(a)}

\Line(250,50)(330,50)
\GlueArc(290,70)(18,-90,270){3}{15}
\Vertex(290,50){1}\Text(290,47)[ct]{$b_1 t_r^2$}
\Text(290,20)[]{(b)}
\end{picture}
\end{center}
\caption{\label{Z_phi one loop}{\small The $t_r^2$ order corrections to the conformal mode.}}
\end{figure}
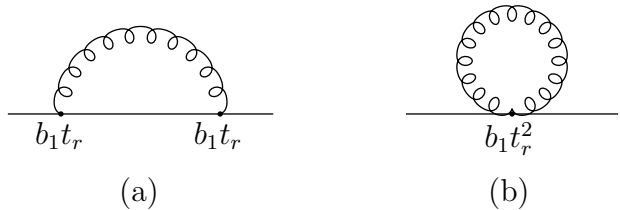

Now, we demonstrate $Z_\phi=1$ in definite calculations. We first consider the two-point function of the conformal mode at the order of $t_r^2$ depicted in Fig.\ref{Z_phi one loop}. From the vertex $L^3_{\phi\phi h}$ in (\ref{Interactions in Wess-Zumino action}), the Feynman diagram (a) in Fig.\ref{Z_phi one loop} gives the contribution\footnote{ 
Although we here introduce the fictitious mass both for the conformal mode and the traceless tensor mode, the result is not changed even if we introduce the mass only for the traceless tensor mode. 
} 
\begin{eqnarray}
    && \int \fr{d^D k}{(2\pi)^D}\phi(k)\phi(-k) \biggl\{ 
       -\fr{b_1}{6} \fr{t_r^2}{(4\pi)^2} \fr{D-2}{2(D-3)}
       \int \fr{d^D l}{(2\pi)^D} \fr{1}{(l^2+z^2)^2 \{(l+k)^2+z^2\}^2}
         \nonumber \\
    && \quad \times
       \biggl[ 6(l^2 k^6+l^6 k^2) + 24 l^4 k^4 -16(l \cdot k)(l^2 k^4+l^4 k^2) 
               -20(l \cdot k)^2 l^2 k^2 
         \nonumber \\
    && \qquad\quad
           -2(l \cdot k)^2(l^4 + k^4) + 8(l \cdot k)^3 (l^2 + k^2) + 8(l \cdot k)^4
         \nonumber \\
    && \qquad\quad
           + \fr{4-D}{3D} \biggl( 
            -36 l^4 k^4 + 24(l \cdot k)(l^2 k^4 + l^4 k^2) + 40(l \cdot k)^2 l^2 k^2
         \nonumber \\
    && \qquad\qquad 
          -4(l \cdot k)^2 (l^4 + k^4) - 16(l \cdot k)^3 (l^2 + k^2) -16 (l \cdot k)^4
         \biggr) \biggr] \biggr\} .
\end{eqnarray}
Carrying out the integral of the momentum $l$ under $z \ll 1$, the inside of the braces $\{~\}$ is calculated as
\begin{equation}
    \fr{2b_1}{(4\pi)^2} k^4 \left[ -3\fr{\tilde{t}_r^2}{(4\pi)^2} \left( \fr{1}{\bar{\eps}} 
                   -\log \fr{z^2}{\mu^2} + \fr{7}{6} \right) \right], 
              \label{Z_phi contribution (a)}
\end{equation}
where $1/\bar{\eps}=1/\eps - \gm + \log 4\pi$. The non-local term $\log (k^2/\mu^2)$ does not appear, which cancels out.

The tadpole diagram (b) in Fig.\ref{Z_phi one loop} comes from the vertex $L^4_{\phi\phi h h}$ in (\ref{Interactions in Wess-Zumino action}). Since the vertices including a derivative of the $h_{\mu\nu}$ field give a vanishing contribution for such a tadpole diagram, only the two terms shown in $L^4_{\phi\phi h h}$ give contributions, which are collected to be
\begin{equation}
    \fr{2b_1}{(4\pi)^2} k^4 \left[ 3 \fr{\tilde{t}_r^2}{(4\pi)^2} \left( \fr{1}{\bar{\eps}} 
                   -\log \fr{z^2}{\mu^2} + \fr{7}{12} \right) \right] .
               \label{Z_phi contribution (b)}
\end{equation}

Combining the contributions (\ref{Z_phi contribution (a)}) and (\ref{Z_phi contribution (b)}), we can see that the UV divergences as well as the IR divergences indeed cancel out. In this way, we can show that $Z_\phi=1$ holds at the order of $t_r^2$. Here, we remark that the sum of the finite terms gives a positive contribution to the coefficient $a_1$ in the effective action (\ref{Riegert action at higher order}).

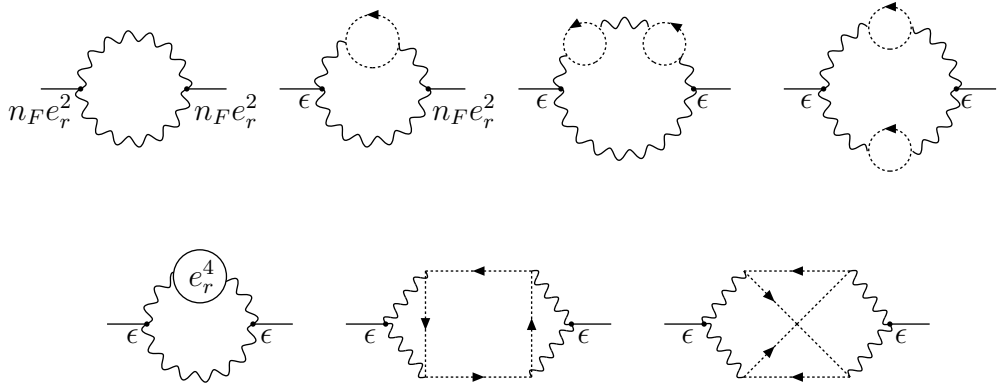
\begin{figure}[h]
\begin{center}
\begin{picture}(380,70)(0,25)
\Line(10,50)(25,50)
\PhotonArc(45,50)(20,0,360){2}{16}
\Line(65,50)(80,50)
\Vertex(25,50){1}\Text(22,48)[rt]{$n_F e_r^2$}
\Vertex(65,50){1}\Text(68,48)[lt]{$n_F e_r^2$}

\Line(100,50)(116,50)
\PhotonArc(135,50)(20,120,60){2}{13}
\DashArrowArc(135,68)(10,-90,270){1}
\Line(156,50)(170,50)
\Vertex(116,50){1}\Text(112,48)[rt]{$\eps$}
\Vertex(156,50){1}\Text(158,48)[lt]{$n_F e_r^2$}

\Line(190,50)(206,50)
\PhotonArc(230,50)(25,150,30){2}{13}
\PhotonArc(230,50)(25,69,111){2}{2.5}
\DashArrowArc(215,67)(8,-60,300){1}
\DashArrowArc(245,67)(8,-120,240){1}
\Line(256,50)(270,50)
\Vertex(206,50){1}\Text(202,48)[rt]{$\eps$}
\Vertex(256,50){1}\Text(258,48)[lt]{$\eps$}

\Line(290,50)(305,50)
\PhotonArc(330,50)(25,-70,70){2}{8}
\PhotonArc(330,50)(25,110,250){2}{8}
\DashArrowArc(330,73)(8,-90,270){1}
\DashArrowArc(330,27)(8,-90,270){1}
\Line(355,50)(370,50)
\Vertex(305,50){1}\Text(302,48)[rt]{$\eps$}
\Vertex(355,50){1}\Text(358,48)[lt]{$\eps$}
\end{picture}
\end{center}
\begin{center}
\begin{picture}(330,70)(0,30)
\Line(10,50)(25,50)
\PhotonArc(45,50)(20,0,360){2}{16}
\GCirc(45,68){10}{1}
\Text(46,68)[]{$e_r^4$}
\Line(65,50)(80,50)
\Vertex(25,50){1}\Text(22,48)[rt]{$\eps$}
\Vertex(65,50){1}\Text(68,48)[lt]{$\eps$}

\Line(100,50)(115,50)
\Photon(115,50)(130,70){2}{4}
\Photon(115,50)(130,30){2}{4}
\DashArrowLine(130,70)(130,30){1}
\DashArrowLine(130,30)(170,30){1}
\DashArrowLine(170,30)(170,70){1}
\DashArrowLine(170,70)(130,70){1}
\Photon(170,70)(185,50){2}{4}
\Photon(170,30)(185,50){2}{4}
\Line(185,50)(200,50)
\Vertex(115,50){1}\Text(112,48)[rt]{$\eps$}
\Vertex(185,50){1}\Text(188,48)[lt]{$\eps$}

\Line(220,50)(235,50)
\Photon(235,50)(250,70){2}{4}
\Photon(235,50)(250,30){2}{4}
\DashArrowLine(250,70)(270,50){1}
        \DashLine(270,50)(290,30){1}
\DashArrowLine(290,30)(250,30){1}
\DashArrowLine(250,30)(270,50){1}
        \DashLine(270,50)(290,70){1}
\DashArrowLine(290,70)(250,70){1}
\Photon(290,70)(305,50){2}{4}
\Photon(290,30)(305,50){2}{4}
\Line(305,50)(320,50)
\Vertex(235,50){1}\Text(232,48)[rt]{$\eps$}
\Vertex(305,50){1}\Text(308,48)[lt]{$\eps$}
\end{picture}
\end{center}
\caption{\label{Z_phi two loop}{\small The $e_r^4$ order corrections to the conformal mode.}}
\end{figure}

Next, we show that the condition $Z_\phi=1$ indeed holds up to the order of $e_r^6$. The loop corrections to the conformal mode at the order of $e_r^2$ are trivially finite. The corrections at the order of $e_r^4$ are depicted in Fig.\ref{Z_phi two loop}. Here, the diagram written by a circle with ``$e_r^4$" inside denotes two-loop diagrams for the ordinary photon self-energy. In the following, for simplicity, diagrams including counterterms inside to subtract UV divergences of subdiagrams are suppressed. At the order of $e_r^4$, there is no counterterm to subtract the overall UV divergence, because the residue of the double pole $b_2$ arises at the order of $e_r^6$ so that the corresponding counterterm to subtract the overall simple pole divergence appears at the order of $e_r^6$, as seen from the Laurent expansion (\ref{Laurent expansion of bG_D}). The sum of the diagrams in Fig.\ref{Z_phi two loop} indeed becomes finite as required.

In the same way, we can demonstrate $Z_\phi=1$ at the order of $e_r^6$ using the Hathrell's results. This result is a consequence of the combination $G_D$ (\ref{G_D}) obtained in Section 2, resulting in the relation (\ref{Hathrell relation}) found by him at the three-loop order.

\subsection{The beta functions}
\noindent

The renormalization factor of the coupling constant $t_r$ has been calculated as
\begin{equation}
    Z_t = 1-  \biggl(\fr{n_F}{80}+\fr{5}{3} \biggr) 
                \fr{{\tilde t}_r^2}{(4\pi)^2}\fr{1}{\eps} 
          - \fr{7 n_F}{288}\fr{{\tilde e}_r^2 {\tilde t}_r^2}{(4\pi)^4}\fr{1}{\eps} 
          + o({\tilde t}_r^4) .
                 \label{Z_t}
\end{equation}
The terms of order of $t^2_r$ is the sum of one-loop contributions from QED \cite{duff} and quantum gravity \cite{ft,amm92} sectors, while the contribution of order of $t_r^2 e_r^2$ comes from two-loop diagrams with no internal line of gravitational fields \cite{dh}.

The beta function of the coupling constant $t_r$ is defined by
\begin{equation}
     \beta_t = \mu \fr{d}{d\mu} {\tilde t}_r .
\end{equation}
The bare quantity defined in the original action should be independent of an arbitrary mass scale $\mu$. This condition yields the equation 
\begin{equation}
   0 = \mu \fr{d}{d\mu} t = \mu \fr{d}{d\mu} (Z_t {\tilde t}_r \mu^\eps),
\end{equation}
and thus the beta function can be written as
\begin{equation}
     \beta_t = -\eps {\tilde t}_r - {\tilde t}_r \fr{\mu}{Z_t} \fr{dZ_t}{d\mu}.
\end{equation}
Noting that the lowest term of the beta function is proportional to $\eps$ such as $\mu d{\tilde t}_r/d\mu=-\eps{\tilde t}_r+ \cdots$ and also $\mu d{\tilde e}_r/d\mu=-\eps{\tilde e}_r +\cdots$, we obtain the beta function
\begin{equation}
      \beta_t = - \biggl( \fr{n_F}{40}+\fr{10}{3} \biggr)
                   \fr{t_r^3}{(4\pi)^2}
                -\fr{7 n_F}{72}\fr{e_r^2 t_r^3}{(4\pi)^4}
                +o(t_r^5) ,
\end{equation}
where we take the limit $\eps \to 0$ after the finite quantity is obtained and remove the tildes on the couplings. Since the beta function becomes negative, the coupling constant for the traceless tensor field indicates the asymptotic freedom.

The renormalization factor for the $U(1)$ gauge field is given by
\begin{eqnarray}
   && Z_3 = 1 -\fr{4n_F}{3}\fr{{\tilde e}_r^2}{(4\pi)^2}\fr{1}{\eps} 
          + \biggl( -2n_F +\fr{8}{27}\fr{n_F^2}{{\tilde b}_1} \biggr) 
               \fr{{\tilde e}_r^4}{(4\pi)^4}\fr{1}{\eps} 
                 \nonumber \\ 
   && \qquad\qquad
          +\biggl( -\fr{8n_F^2}{9} + \fr{8}{81}\fr{n_F^3}{{\tilde b}_1} \biggr) 
               \fr{{\tilde e}_r^6}{(4\pi)^6}\fr{1}{\eps^2} 
          + o({\tilde e}_r^2 {\tilde t}_r^2, {\tilde t}_r^4). 
                \label{Z_3}
\end{eqnarray}
Here, the corrections proportional to $1/b_1$ are the contributions from diagrams with an internal line of the conformal mode that arise at the order of $e_r^4$. The results of double pole divergences that arise at the order of $e_r^6$ are added.  There is no contribution of order of $t_r^2$, because the UV divergences from diagrams with an internal line of the traceless tensor field cancel out at this order.

The beta function of the QED coupling constant is defined by
\begin{equation}
     \beta_e = \mu \fr{d{\tilde e}_r}{d\mu} .
         \label{beta_e function 1} 
\end{equation}
Since the Ward-Takahashi identity $Z_1=Z_2$ holds even if quantum gravity is coupled, the renormalization factor of the coupling constant is given by $Z_e=Z_3^{-1/2}$. Thus, the beta function can be written as
\begin{equation}
     \beta_e =-\eps {\tilde e}_r + \fr{{\tilde e}_r}{2} \fr{\mu}{Z_3} \fr{dZ_3}{d\mu} .
         \label{beta_e function 2}
\end{equation}
Since the bare quantity $b$ is independent of the arbitrary mass scale $\mu$, the dimensionless constant defined by (\ref{scale dependence of coupling constant}) satisfies the equation $\mu d{\tilde b}_1^{-1}/d\mu = -2\eps {\tilde b}_1^{-1}$. 
We here regard the constant ${\tilde b}_1$ as an arbitrary constant when we compute the beta function. After a finite expression is obtained, we substitute the definite value. This treatment is consistent with the result of $Z_3$ (\ref{Z_3}) reflected in the relations between the residues of simple poles and double poles. In this way, we obtain the finite expression of the beta function,
\begin{equation}
    \beta_e = \fr{4 n_F}{3}\fr{e_r^3}{(4\pi)^2}
           + \biggl( 4 n_F - \fr{8}{9}\fr{n_F^2}{b_1}
                \biggr) \fr{e_r^5}{(4\pi)^4}
           + o(e_r^3t_r^2) .
\end{equation}
We here remark that the effect of quantized gravity gives a negative contribution. Substituting the value of $b_1$ (\ref{b_1 and b_2}), the whole of the $e_r^5$ order term becomes negative if $n_F \geq 24$.


\section{Renormalization of Composite Fields}
\setcounter{equation}{0}
\noindent

In this section, we discuss renormalization of the cosmological constant term and the Einstein action, which are given by composite fields with exponential factor of the conformal mode even at the vanishing limit of the coupling constant.

\subsection{Vertices in lower-derivative actions}
\noindent

The cosmological constant term is simply written in terms of the exponential factor of the conformal mode as
\begin{equation}
    I_\Lam = \Lam \int d^D x \sq{g} = \Lam \int d^D x  e^{D\phi},
\end{equation}
while the Einstein action has the form expanded in the coupling constant as
\begin{eqnarray}
    I_{\rm EH} &=& -\fr{M_\P^2}{2} \int d^D x \sq{g} R 
               \nonumber \\
            &=&  -\fr{M_\P^2}{2} \int d^D x e^{(D-2)\phi} \Bigl\{ \bR -2(D-1)\bnb^2 \phi
                     -(D-1)(D-2) \bnb_\lam \phi \bnb^\lam \phi  \Bigr\}
                \nonumber \\
            &=& \fr{3}{2}\fr{D-1}{3}M_\P^2 \int d^D x e^{(D-2)\phi} \biggl\{ 
                  \pd^2 \phi + \fr{D-2}{D-1} t h_{\mu\nu} \left( -\pd_\mu \pd_\nu \phi +\pd_\mu \phi \pd_\nu \phi \right)
                     \nonumber \\
            && \qquad\quad
               + \fr{t^2}{2} h_{\mu\lam}h_{\nu\lam} \pd_\mu \pd_\nu \phi 
               + \fr{t^2}{2} h_{\mu\nu} \pd_\mu h_{\nu\lam} \pd_\lam \phi
               -\fr{D-3}{2(D-1)} t^2 h_{\mu\nu} \pd_\lam h_{\lam\mu} \pd_\nu \phi 
                    \nonumber \\
           && \qquad\quad
               + \fr{t^2}{4(D-1)} \pd_\lam h_{\mu\nu} \pd_\lam h_{\mu\nu}
               -\fr{t^2}{2(D-1)} \pd_\mu h_{\mu\lam} \pd_\nu h_{\nu\lam} + o(t^3) 
                  \biggr\} .
\end{eqnarray}
These terms are renormalized by redefining the bare cosmological constant and the square of the bare Planck mass as
\begin{eqnarray}
    \Lam &=& Z_\Lam \Lam_r,
         \nonumber \\
    \fr{D-1}{3} M_\P^2 &=& Z_{\rm EH} M_\P^{r 2} ,
\end{eqnarray}
respectively. Here $\Lam_r$ and $M_\P^r$ are the renormalized mass scales. Those with canonical dimensions denoted by symbols with tildes are defined as
\begin{equation}
     \Lam_r = \tilde{\Lam}_r \mu^{-2\eps}, \qquad M_\P^{r 2} = \tilde{M}_\P^{r 2} \mu^{-2\eps}.
     \label{canonical mass scales}
\end{equation}
The renormalization factors are expanded as
\begin{eqnarray}
   Z_\Lam &=& 1 + \fr{u_1}{D-4} + \fr{u_2}{(D-4)^2} + \cdots,
       \nonumber \\
   Z_{\rm EH} &=& 1 + \fr{v_1}{D-4} + \fr{v_2}{(D-4)^2} + \cdots.
\end{eqnarray}

Since the conformal mode is not renormalized, the cosmological constant is expanded using only the Laurent expansion of $Z_\Lam$ in the form
\begin{eqnarray}
    I_\Lam &=& \Lam_r \int d^D x \biggl\{ 
               \left( 1+ \fr{u_1}{D-4} + \fr{u_2}{(D-4)^2} + \cdots \right) e^{4\phi}
                  \nonumber \\
          &&  \qquad\qquad\quad
               +\left( D-4 + u_1 + \fr{u_2}{D-4} + \cdots \right) \phi e^{4\phi}
                  \nonumber \\
          && \qquad\qquad\quad
               + \half \left( (D-4)^2 +(D-4)u_1 + u_2 +\cdots \right) \phi^2 e^{4\phi}
                  \nonumber \\
          && \qquad\qquad\quad
               + \cdots \biggr\} .
             \label{Laurent expansion of cosmological constant}
\end{eqnarray}
The Einstein action is expanded in terms of the renormalized quantities using the renormalization factor $Z_{\rm EH}$ and those for the coupling constant and traceless tensor fields as
\begin{eqnarray}
    I_{\rm EH} &=& \fr{3}{2}M_\P^{r2} \int d^D x  \biggl\{ 
                   \left( 1 + \fr{v_1}{D-4} +\cdots \right) 
                   e^{2\phi} \biggl( \pd^2 \phi -\fr{2}{3} t_r h^r_{\mu\nu} \pd_\mu \pd_\nu \phi
                      \nonumber \\
               && \qquad\qquad\qquad\qquad\qquad
                   + \fr{2}{3} t_r h^r_{\mu\nu} \pd_\mu  \phi \pd_\nu \phi
                   + \fr{t_r^2}{2} h^r_{\mu\lam}h^r_{\nu\lam} \pd_\mu \pd_\nu \phi  + \cdots \biggr)
                      \nonumber \\
               && \qquad\qquad\qquad
                   + \left(D-4 + v_1 +\cdots \right) e^{2\phi} 
                      \biggl( \phi \pd^2 \phi - \fr{2}{3}t_r \phi \pd_\mu \pd_\nu \phi h^r_{\mu\nu}
                        \nonumber \\
               && \qquad\qquad\qquad\qquad\qquad
                                   -\fr{1}{9}t_r  \pd_\mu \pd_\nu \phi h^r_{\mu\nu}
                                     + \cdots \biggr) 
                      \nonumber \\
               && \qquad\qquad\qquad
                   + \cdots \biggr\} .
\end{eqnarray}

\subsection{Scaling dimension of the cosmological constant}
\noindent

We first consider the renormalization of the cosmological constant. In the following, we only consider the gravitational interactions and do not take care of the matter contents, which are represented by the constant $b_1$. The computation is carried out in the large $b_1$ limit, which corresponds to a large number limit of matter fields. Furthermore, in this section, we take care only the UV divergences to compute anomalous dimensions, while the IR divergence will be discussed when we compute the effective cosmological constant in Section 6.

\begin{figure}[h]
\begin{center}
\begin{picture}(300,110)(0,-10)
\CArc(50,50)(20,0,360)
\Vertex(50,30){1}
\Text(50,7)[]{$\underbrace{~\cdots~}$}\Text(50,-3)[]{$n$}
\Line(38,10)(50,30)
\Line(62,10)(50,30)
\Text(50,-20)[]{(a)}

\CArc(150,50)(20,0,360)
\Line(150,30)(150,70) 
   \Vertex(150,70){1}\Text(150,75)[b]{$\eps b_1$}
\Vertex(150,30){1}
\Text(150,7)[]{$\underbrace{~\cdots~}$}\Text(150,-3)[]{$n$}
\Line(138,10)(150,30)
\Line(162,10)(150,30)
\Text(150,-20)[]{(b)}

\CArc(250,55)(25,0,360)
\Oval(250,50)(20,15)(0)
\Vertex(250,30){1}
\Text(250,7)[]{$\underbrace{~\cdots~}$}\Text(250,-3)[]{$n$}
\Line(238,10)(250,30)
\Line(262,10)(250,30)
\Text(250,-20)[]{(c)}
\end{picture} 
\end{center}
\caption{\label{Z_Lambda 1}{\small The $1/b_1$ and $1/b_1^2$ order corrections to the cosmological constant.}}
\end{figure}
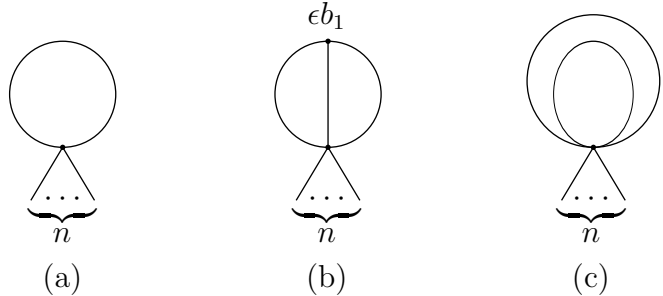

The Feynman diagrams contributing to anomalous dimensions at the first and second orders of the expansion in $1/b_1$ are depicted in Fig.\ref{Z_Lambda 1}. Here, we carry out computations by expanding the exponential factor as $e^{4\phi}=\sum_n (4 \phi)^n/n!$. The $\phi^3$ vertex with $\eps b_1$ in diagram (b) is the induced vertex from the Laurent expansion of the $G_D$ action (\ref{Laurent expansion of bG_D}). The renormalization factor to subtract the UV divergences from these diagrams is given by
\begin{equation}
   Z_\Lam = 1 - \fr{2}{{\tilde b}_1} \fr{1}{\eps} -\fr{2}{{\tilde b}_1^2} \fr{1}{\eps} 
             + \fr{2}{{\tilde b}_1^2}\fr{1}{\eps^2} +\cdots .
             \label{Z_Lam contribution}
\end{equation}
The simple pole terms come from diagrams (a) and (b), respectively. The double pole term comes from diagram (c) with separate loops, where counterterm diagrams to subtract subdivergences are suppressed. Thus, we obtain the residues of the renormalization factor to be $u_1=4/{\tilde b}_1+4/{\tilde b}_1^2$ and $u_2=8/{\tilde b}_1^2$.

\begin{figure}[h]
\begin{center}
\begin{picture}(300,120)(0,-10)
\CArc(50,50)(20,0,360)
\Vertex(50,30){1}
\Text(50,7)[]{$\underbrace{~\cdots~}$}\Text(50,-3)[]{$n$}
\Line(30,30)(50,30)\Text(23,30)[]{$u_1$}
\Line(38,10)(50,30)
\Line(62,10)(50,30)
\Text(50,-20)[]{(a)}

\CArc(150,55)(25,0,360)
\Oval(150,50)(20,15)(0)
\Vertex(150,30){1}
\Text(150,7)[]{$\underbrace{~\cdots~}$}\Text(150,-3)[]{$n$}
\Line(125,30)(150,30)\Text(120,30)[]{$\eps$}
\Line(138,10)(150,30)
\Line(162,10)(150,30)
\Text(150,-20)[]{(b)}

\CArc(250,55)(25,0,360)
\Oval(250,50)(20,15)(0)
\Vertex(250,30){1}
\Text(250,7)[]{$\underbrace{~\cdots~}$}\Text(250,-3)[]{$n$}
\Line(250,80)(250,100)\Vertex(250,80){1}\Text(263,85)[]{$\eps b_1$}
\Line(238,10)(250,30)
\Line(262,10)(250,30)
\Text(250,-20)[]{(c)}
\end{picture} 
\end{center}
\caption{\label{Z_Lambda 1(induced)}{\small The $1/b_1^2$ order corrections to the induced $\phi e^{4\phi}$ vertex.}}
\end{figure}
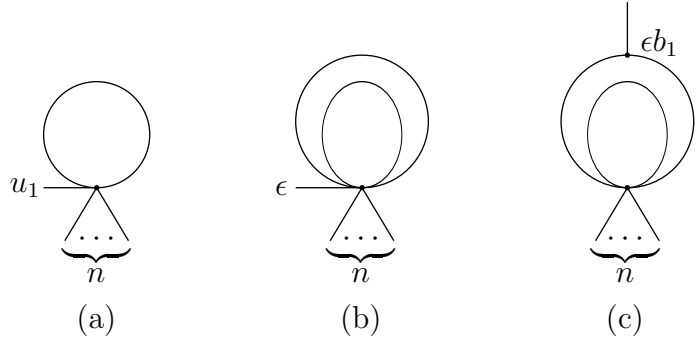

The potentially divergent diagrams proportional to the induced vertex of $\phi e^{4\phi}$ are given at the order of $1/b_1^2$, which are depicted in Fig.\ref{Z_Lambda 1(induced)}. The first two diagrams, (a) and (b), are constructed from the induced vertices given in the second line of the Laurent expansion (\ref{Laurent expansion of cosmological constant}). The sum of the UV divergences from these diagrams, however, exactly cancels the simple-pole counterterm of $\phi e^{4\phi}$ induced by the residue $u_2$. The diagram (c) also gives a finite contribution because the UV divergence cancels that from its associate counterterm diagram to subtract subdivergences, not depicted here. Thus, these diagrams do not contribute to the renormalization factor. The UV divergences proportional to the induced vertices will be renormalized by using the information of the residues $u_n$.

The anomalous dimension of the cosmological constant is defined by
\begin{equation}
     \gm_\Lam = -\fr{\mu}{\tilde{\Lam}_r} \fr{d \tilde{\Lam}_r}{d\mu} .
\end{equation}
Using equation (\ref{canonical mass scales}) and the fact that the bare cosmological constant satisfies the equation $d\Lam/d\mu=0$, we obtain
\begin{equation}
    \gm_\Lam = -2 \eps + \fr{\mu}{Z_\Lam}\fr{d Z_\Lam}{d \mu}
             = \fr{4}{b_1} + \fr{8}{b_1^2} + \cdots 
           \label{annomalous dim. of Lambda}
\end{equation}
in the large $b_1$ expansion.

This result can be compared with the exact expression computed using the conformal algebra. The anomalous dimension represents a response to a scale transformation, or conformal transformation, and thus the conformal-mode dependence of the renormalized cosmological constant is defined by $\dl_\phi L_\Lam =(4+\gm_\Lam)L_\Lam$, where $4$ is the canonical value. The conformal-mode dependence obtained by solving the conformal invariance condition is given by $e^{\gm_0 \phi}$ with $\gm_0=2b_1 ( 1-\sq{1-4/b_1} )$, and thus we obtain the relationship $\gm_0= 4+\gm_\Lam$. Hence, the exact solution of $\gm_\Lam$ is given by
\begin{equation}
    \gm_\Lam = \gm_0-4 = \fr{4}{b_1}+\fr{8}{b_1^2}+\fr{20}{b_1^3}+\cdots .
\end{equation}
The first two terms agree with the results in (\ref{annomalous dim. of Lambda}).

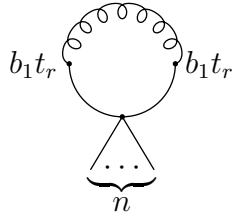
\begin{figure}[h]
\begin{center}
\begin{picture}(200,80)(-50,10)
\GlueArc(50,50)(20,0,180){3}{7}
\CArc(50,50)(20,180,360)
\Vertex(30,50){1}\Vertex(70,50){1}
\Text(26,50)[r]{$b_1 t_r$}\Text(74,50)[l]{$b_1 t_r$}
\Vertex(50,30){1}
\Text(50,7)[]{$\underbrace{~\cdots~}$}\Text(50,-3)[]{$n$}
\Line(38,10)(50,30)
\Line(62,10)(50,30)
\end{picture} 
\end{center}
\caption{\label{Z_Lambda 2}{\small The $t_r^2$ order correction to the cosmological constant.}}
\end{figure}

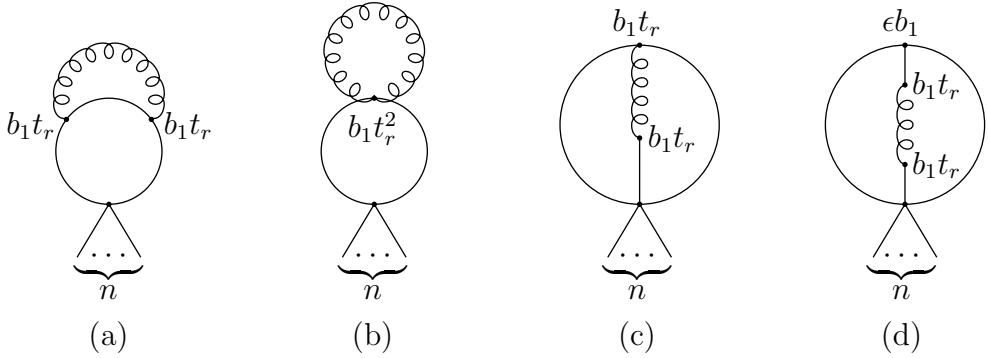
\begin{figure}[h]
\begin{center}
\begin{picture}(400,120)(0,-10)
\GlueArc(50,70)(18,-25,205){3}{9}
\Vertex(66,62){1}\Vertex(34,62){1}
\Text(30,60)[r]{$b_1 t_r$}\Text(71,60)[l]{$b_1 t_r$}
\CArc(50,50)(20,0,360)
\Vertex(50,30){1}
\Text(50,7)[]{$\underbrace{~\cdots~}$}\Text(50,-3)[]{$n$}
\Line(38,10)(50,30)
\Line(62,10)(50,30)
\Text(50,-20)[]{(a)}

\GlueArc(150,87)(16,-90,270){3}{12}
\Vertex(150,70){1}
\Text(150,65)[t]{$b_1 t_r^2$}
\CArc(150,50)(20,0,360)
\Vertex(150,30){1}
\Text(150,7)[]{$\underbrace{~\cdots~}$}\Text(150,-3)[]{$n$}
\Line(138,10)(150,30)
\Line(162,10)(150,30)
\Text(150,-20)[]{(b)}

\CArc(250,60)(30,0,360)
\Gluon(250,55)(250,90){3}{4} 
   \Vertex(250,90){1}\Text(250,94)[b]{$b_1 t_r$}
   \Vertex(250,55){1}\Text(254,55)[l]{$b_1 t_r$}
\Line(250,30)(250,55)
\Vertex(250,30){1}
\Text(250,7)[]{$\underbrace{~\cdots~}$}\Text(250,-3)[]{$n$}
\Line(238,10)(250,30)
\Line(262,10)(250,30)
\Text(250,-20)[]{(c)}

\CArc(350,60)(30,0,360) 
\Line(350,75)(350,90) 
   \Vertex(350,90){1}\Text(350,94)[b]{$\eps b_1$}
\Gluon(350,45)(350,75){3}{3}
   \Vertex(350,75){1}\Text(354,75)[l]{$b_1 t_r$}
   \Vertex(350,45){1}\Text(354,45)[l]{$b_1 t_r$}
\Line(350,30)(350,45)
\Vertex(350,30){1}
\Text(350,7)[]{$\underbrace{~\cdots~}$}\Text(350,-3)[]{$n$}
\Line(338,10)(350,30)
\Line(362,10)(350,30)
\Text(350,-20)[]{(d)}
\end{picture} 
\end{center}
\caption{\label{Z_Lambda 3}{\small The $t_r^2/b_1$ order corrections to the cosmological constant.}}
\end{figure}

Furthermore, there are corrections from the Feynman diagrams including the interactions with the traceless tensor mode given in (\ref{Interactions in Wess-Zumino action}). The $o(t_r^2)$ and $o(t_r^2/b_1)$ corrections are given by the diagrams in Fig.\ref{Z_Lambda 2} and Fig.\ref{Z_Lambda 3}, respectively. The sum of these diagrams produces only the simple-pole divergence.  The corresponding part of the renormalization factor is computed as
\begin{eqnarray}
    Z_\Lam |_{t_r^2} &=& - \fr{1}{6} \fr{\tilde{t}_r^2}{(4\pi)^2} \fr{1}{\eps} 
                            + \left( -\fr{7}{2} + \fr{1}{3} -\fr{1}{2}  \right) 
                               \fr{1}{\tilde{b}_1} \fr{\tilde{t}_r^2}{(4\pi)^2} \fr{1}{\eps}
               \nonumber \\ 
        &=&  -  \left( \fr{1}{6} + \fr{11}{3}\fr{1}{{\tilde b}_1} \right) 
                             \fr{{\tilde t}_r^2}{(4\pi)^2} \fr{1}{\eps} .
                    \label{Z_Lam contribution at order t_r^2}
\end{eqnarray}
Here, the first term of the $o(t_r^2/b_1)$ divergences in the first line comes from the sum of the contributions from diagrams (a) and (b) in Fig.\ref{Z_Lambda 3}, which is computed using the results given in Section 4.4, while the second and third terms are contributions from diagrams (c) and (d), respectively.

\begin{figure}[h]
\begin{center}
\begin{picture}(250,120)(0,-10)
\Text(18,50)[]{$\sum_{n^\pp}$}
\Vertex(50,70){1}
\Text(50,93)[]{$\overbrace{~\cdots~}$}\Text(50,103)[]{$n-n^\pp$}
\Line(38,90)(50,70)
\Line(62,90)(50,70)
\CArc(50,50)(20,0,360)
\Vertex(50,30){1}
\Text(50,7)[]{$\underbrace{~\cdots~}$}\Text(50,-3)[]{$n^\pp$}
\Line(38,10)(50,30)
\Line(62,10)(50,30)
\Text(50,-20)[]{(a)}

\Text(168,50)[]{$\sum_{n^\pp}$}
\Vertex(200,70){1}
\Text(200,93)[]{$\overbrace{~\cdots~}$}\Text(200,103)[]{$n-n^\pp$}
\Line(188,90)(200,70)
\Line(212,90)(200,70)
\CArc(200,50)(20,90,270)
\Text(194,74)[r]{$t_r$}\Text(194,27)[r]{$t_r$}
\GlueArc(200,50)(20,-90,90){3}{7}
\Vertex(200,30){1}
\Text(200,7)[]{$\underbrace{~\cdots~}$}\Text(200,-3)[]{$n^\pp$}
\Line(188,10)(200,30)
\Line(212,10)(200,30)
\Text(200,-20)[]{(b)}
\end{picture} 
\end{center}
\caption{\label{Z_Lambda 4}{\small The $M_\P^{r 4}/b_1^2$ and $M_\P^{r 4}t_r^2/b_1$ order corrections to the cosmological constant.}}
\end{figure}
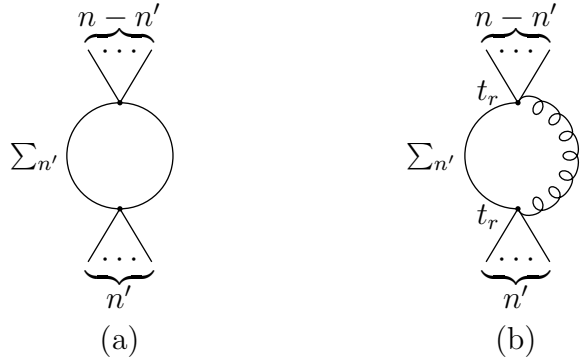

When the Einstein action is coupled, there are loop corrections dependent on the mass scales. They are given by diagrams (a) and (b) in Fig.\ref{Z_Lambda 4} up to the order of $t_r^2$. The corresponding part of the renormalization factor is given by 
\begin{eqnarray}
    Z_\Lam |_{\rm mass-dep.} = \left( \fr{6}{{\tilde b}_1^2} 
                    + \fr{1}{{\tilde b}_1}\fr{{\tilde t}_r^2}{(4\pi)^2} \right) 
                       \fr{3\pi^2{\tilde M}_\P^{r 4}}{2{\tilde \Lam}_r} \fr{1}{\eps} +\cdots.
                 \label{Z_Lam contribution dependent on mass}
\end{eqnarray}

Adding (\ref{Z_Lam contribution at order t_r^2}) and (\ref{Z_Lam contribution dependent on mass}) to the renormalization factor (\ref{Z_Lam contribution}), we obtain the following anomalous dimension:
\begin{equation}
   \gm_\Lam = \fr{4}{b_1} + \fr{8}{b_1^2} 
            + \left( \fr{1}{3} + \fr{44}{3}\fr{1}{b_1} \right) \fr{t_r^2}{(4\pi)^2}
            -  \left( \fr{6}{b_1^2} + \fr{1}{b_1}\fr{t_r^2}{(4\pi)^2} \right) 
                  \fr{3\pi^2 M_\P^{r 4}}{\Lam_r} +  \cdots
\end{equation}
in the large $b_1$ expansion.

\subsection{Scaling dimension of the Planck mass}
\noindent

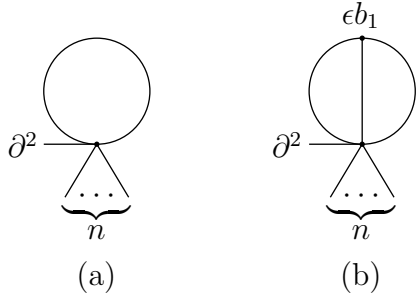
\begin{figure}[h]
\begin{center}
\begin{picture}(200,100)(0,-10)
\CArc(50,50)(20,0,360)
\Vertex(50,30){1}
\Text(50,7)[]{$\underbrace{~\cdots~}$}\Text(50,-3)[]{$n$}
\Line(38,10)(50,30)
\Line(62,10)(50,30)
\Line(30,30)(50,30)\Text(28,30)[r]{$\pd^2$}
\Text(50,-20)[]{(a)}

\CArc(150,50)(20,0,360)
\Line(150,30)(150,70) 
   \Vertex(150,70){1}\Text(150,74)[b]{$\eps b_1$}
\Vertex(150,30){1}
\Text(150,7)[]{$\underbrace{~\cdots~}$}\Text(150,-3)[]{$n$}
\Line(138,10)(150,30)
\Line(162,10)(150,30)
\Line(130,30)(150,30)\Text(128,30)[r]{$\pd^2$}
\Text(150,-20)[]{(b)}
\end{picture} 
\end{center}
\caption{\label{Z_EH 1}{\small The $1/b_1$ and $1/b_1^2$ order corrections to the Planck mass.}}
\end{figure}

\begin{figure}[h]
\begin{center}
\begin{picture}(400,120)(0,-10)
\GlueArc(50,50)(20,0,180){3}{7}
\Vertex(30,50){1}\Vertex(70,50){1}
\Text(26,50)[r]{$b_1 t_r$}\Text(74,50)[l]{$b_1 t_r$}
\CArc(50,50)(20,180,360)
\Vertex(50,30){1}
\Text(50,7)[]{$\underbrace{~\cdots~}$}\Text(50,-3)[]{$n$}
\Line(38,10)(50,30)
\Line(62,10)(50,30)
\Line(30,30)(50,30)\Text(28,30)[r]{$\pd^2$}
\Text(50,-20)[]{(a)}

\GlueArc(130,50)(20,-90,270){3}{14}
\Text(130,33)[b]{$t_r^2$}
\Vertex(130,30){1}
\Text(130,7)[]{$\underbrace{~\cdots~}$}\Text(130,-3)[]{$n$}
\Line(118,10)(130,30)
\Line(142,10)(130,30)
\Line(110,30)(130,30)\Text(108,30)[r]{$\pd^2$}
\Text(130,-20)[]{(b)}

\GlueArc(210,50)(20,-90,90){3}{7}
\Vertex(210,70){1}
\Text(210,75)[b]{$b_1 t_r$}
\Text(210,33)[b]{$t_r$}
\CArc(210,50)(20,90,270)
\Vertex(210,30){1}
\Text(210,7)[]{$\underbrace{~\cdots~}$}\Text(210,-3)[]{$n$}
\Line(198,10)(210,30)
\Line(222,10)(210,30)
\Line(190,30)(210,30)\Text(188,30)[r]{$\pd^2$}
\Text(210,-20)[]{(c)}

\GlueArc(290,50)(20,-90,90){3}{7}
\Line(290,70)(290,90) 
   \Vertex(290,70){1}\Text(280,73)[b]{$b_1 t_r$}
   \Text(290,94)[b]{$\pd^2$}
\Text(290,33)[b]{$t_r$}
\CArc(290,50)(20,90,270)
\Vertex(290,30){1}
\Text(290,7)[]{$\underbrace{~\cdots~}$}\Text(290,-3)[]{$n$}
\Line(278,10)(290,30)
\Line(302,10)(290,30)
\Text(290,-20)[]{(d)}

\GlueArc(370,50)(20,0,90){3}{4}
\Line(370,70)(370,90) 
   \Vertex(370,70){1}\Text(360,73)[b]{$b_1 t_r$}
   \Text(370,94)[b]{$\pd^2$}
\Text(380,50)[]{$b_1t_r$}
\CArc(370,50)(20,90,360)
\Vertex(370,30){1}\Vertex(390,50){1}
\Text(370,7)[]{$\underbrace{~\cdots~}$}\Text(370,-3)[]{$n$}
\Line(358,10)(370,30)
\Line(382,10)(370,30)
\Text(370,-20)[]{(e)}
\end{picture} 
\end{center}
\caption{\label{Z_EH 2}{\small The $t_r^2$ order corrections to the Planck mass.}}
\end{figure}
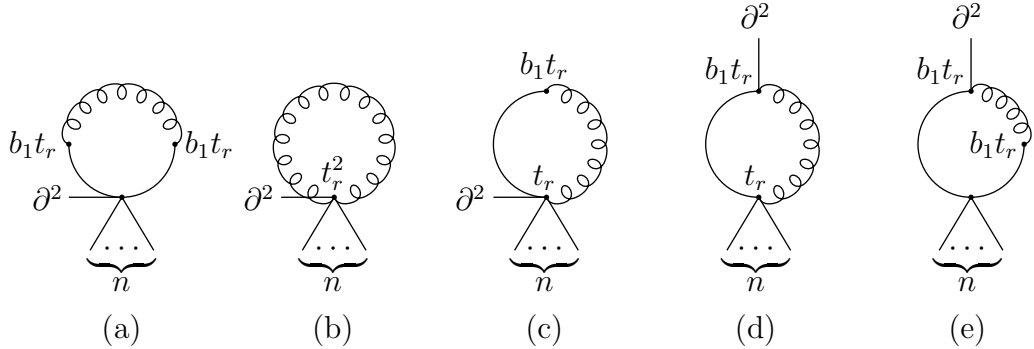

Next, we consider corrections to the Einstein action up to the order of $t_r^2$. The Feynman diagrams that yield simple poles are depicted in Fig.\ref{Z_EH 1} and Fig.\ref{Z_EH 2}. The renormalization factor is computed as
\begin{eqnarray}
   Z_\EH-1 &=& -\fr{1}{2{\tilde b}_1} \fr{1}{\eps} - \fr{1}{4{\tilde b}_1^2} \fr{1}{\eps} 
                  + \left( -\fr{1}{24} -\fr{9}{8} +\fr{2}{3} -\fr{1}{12} \right) 
                      \fr{{\tilde t}_r^2}{(4\pi)^2}\fr{1}{\eps} + \cdots
              \nonumber \\
      &=& -\fr{1}{2{\tilde b}_1} \fr{1}{\eps} - \fr{1}{4{\tilde b}_1^2} \fr{1}{\eps} 
                  -  \fr{7}{12} \fr{{\tilde t}_r^2}{(4\pi)^2}\fr{1}{\eps} + \cdots.
\end{eqnarray}
The contributions of orders $1/b_1$ and $1/b_1^2$ come from diagrams (a) and (b) in Fig.\ref{Z_EH 1}, respectively. The $t_r^2$ order contributions in the first line are, from the left, given by diagrams (a), (b), (d) and (e) in Fig.\ref{Z_EH 2}, respectively, while diagram (c) becomes finite.

The anomalous dimension for the square of the Planck mass is defined by
\begin{equation}
    \gm_\EH = - \fr{\mu}{\tilde{M}_\P^{r 2}} \fr{d \tilde{M}_\P^{r 2}}{d\mu}. 
\end{equation}
Using the equation for the bare Planck mass $dM_\P^2/d\mu=0$, we obtain
\begin{eqnarray}
    \gm_\EH &=& -2 \eps + \fr{\mu}{Z_\EH}\fr{d Z_\EH}{d \mu}
                \nonumber \\
            &=& \fr{1}{b_1} + \fr{1}{b_1^2} + \fr{7}{6}\fr{t_r^2}{(4\pi)^2} + \cdots.
         \label{anomalous dimension of planck mass}
\end{eqnarray}

Here, we compare this result with the exact expression derived from CFT, which is given by 
\begin{equation}
    \gm_\EH = 2b_1 \left( 1-\sq{1-\fr{2}{b_1}} \right) -2
        = \fr{1}{b_1}+\fr{1}{b_1^2}+\fr{5}{4b_1^3}+\cdots .
\end{equation}
This agrees with (\ref{anomalous dimension of planck mass}) in the CFT limit of $t_r \to 0$.

\section{Renormalization Group and Effective Cosmological Constant}
\setcounter{equation}{0}
\noindent

We here study the renormalization group equation \cite{gl,symanzik,callan,thooft,weinberg73} for the effective action, especially for the effective cosmological constant, in which the scale parameter will be identified with the constant mode of the $\phi$ field.

\subsection{The {}$^\pp$t Hooft-Weinberg equation}
\noindent

Consider the renormalized $n$-point Green functions $\Gm_r^{(n)}$ of composite fields of the conformal mode with canonical mass dimension $d_0$, like $\sq{g}$ and $\sq{g}R$ with $d_0=0$ and $2$, respectively.  Since the conformal mode is not renormalized such that $Z_\phi=1$, the unrenormalized Green function $\Gm^{(n)}$ is the same to the renormalized one. Therefore, the renormalization group equation is represented as $d \Gm^{(n)}/d\mu= d \Gm_r^{(n)}/d\mu=0$. By using the chain rule for differentiation we have 
\begin{eqnarray}
   && \biggl( \mu \fr{\pd}{\pd\mu} + \b_t(t_r)\fr{\pd}{\pd t_r} 
          - \gm_\Lam \left( t_r,\Lam_r, M_\P^{r 2} \right) \Lam_r \fr{\pd}{\pd \Lam_r}
            \nonumber \\
   && \qquad\quad
          - \gm_\EH \left( t_r \right) M_\P^{r 2} \fr{\pd}{\pd M_\P^{r 2}}
         \biggr) \Gm_r^{(n)} \left( \lam k,t_r, \Lam_r, M_\P^{r 2}, \mu \right) =0,
         \label{renormalization group equation 1}
\end{eqnarray} 
where $k$ denotes a set of external momenta on the flat background and we introduced the dimensionless scale parameter $\lam$. The renormalization group quantities $\b_t$, $\gm_\Lam$ and $\gm_\EH$ have been defined in the previous sections. We here suppressed the tildes on the dimensionless quantities such as $\tildet_r$ as no confusion will be occurred.

Based on naive dimension counting, we find that the Green function has the following form:
\begin{equation}
    \Gm_r^{(n)} \left( \lam k, t_r, \Lam_r, M_\P^{r 2} \right) 
     = \mu^{4-nd_0} \Om^{(n)} \left( \fr{\lam k}{\mu}, t_r, \fr{\Lam_r}{\mu^4}, \fr{M_\P^{r 2}}{\mu^2} \right) ,
\end{equation} 
where $\Om^{(n)}$ is the dimensionless function. Taking into account the above general form, we obtain an identity
\begin{equation}
   \left(  \mu \fr{\pd}{\pd \mu} + \lam \fr{\pd}{\pd \lam} 
      + 4 \Lam_r \fr{\pd}{\pd \Lam_r}   + 2 M_\P^{r 2} \fr{\pd}{\pd M_\P^{r 2}} 
      -4 + nd_0 \right) \Gm_r^{(n)} =0 .
        \label{identity equation}
\end{equation}
Combining equations (\ref{renormalization group equation 1}) and (\ref{identity equation}), we obtain the following equation:
\begin{eqnarray}
   && \biggl(  -\lam \fr{\pd}{\pd\lam} + \b_t(t_r)\fr{\pd}{\pd t_r} 
          - \left[ 4+ \gm_\Lam \left( t_r,\Lam_r, M_\P^{r 2} \right) 
                    \right] \Lam_r \fr{\pd}{\pd \Lam_r}
            \nonumber \\
   && \quad
          - \left[ 2 + \gm_\EH \left( t_r \right) 
                   \right] M_\P^{r 2} \fr{\pd}{\pd M_\P^{r 2}}
          + 4 -nd_0 
         \biggr) \Gm_r^{(n)} \left( \lam k, t_r, \Lam_r, M_\P^{r 2},\mu \right) =0.
           \nonumber \\
   && \label{renormalization group equation 2}
\end{eqnarray}

Here, we set the scale parameter to be
\begin{equation}
     \lam = e^\s
\end{equation}
and introduce the running coupling constant $\overline{t}_r(\s)$ and the running mass scales $\overline{\Lam}(\s)$ and $\overline{M}_\P(\s)$ defined by the equations
\begin{eqnarray}
  -\fr{d}{d\s}\overline{t}_r &=& \b_t \left( \overline{t}_r \right),
              \nonumber \\
  -\fr{d}{d\s} \overline{\Lam} &=& 
       -\left[ 4+\gm_\Lam \left( \overline{t}_r, \overline{\Lam},\overline{M}_\P^2 \right) 
                                    \right] \overline{\Lam} ,
              \nonumber \\
  -\fr{d}{d\s}\overline{M}^2_\P  &=& 
      -\left[ 2+\gm_\EH \left( \overline{t}_r \right) \right] \overline{M}^2_\P .
       \label{running mass equation}
\end{eqnarray}
If we replace $t_r$, $\Lam_r$ and $M_\P^{r 2}$ in equation (\ref{renormalization group equation 2}) by the running coupling constant $\overline{t}_r(\s)$ and the running mass scales $\overline{\Lam}(\s)$ and $\overline{M}_\P^2(\s)$, respectively, we find that with the help of the equations defining these quantities, this equation is transformed into a total derivative
\begin{equation}
   \left( -\fr{d}{d\s} + 4-nd_0 \right) \Gm_r^{(n)} 
   \left( e^\s k,\overline{t}_r(\s), \overline{\Lam}(\s), \overline{M}_\P^2(\s),\mu \right) =0 .
\end{equation}
Thus, we obtain
\begin{equation}
    \Gm_r^{(n)} \left( e^\s k,\overline{t}_r(\s), \overline{\Lam}(\s), \overline{M}_\P^2(\s),\mu \right) 
    = \Gm_r^{(n)} \left( k, t_r, \Lam_r, M_\P^{r 2},\mu \right) 
         e^{(4-nd_0)\s},
         \label{solution of tHooft-Weinberg equation}
\end{equation}
where we take the conditions
\begin{equation}
    \overline{t}_r(\s=0)=t_r, \quad \overline{\Lam} (\s=0) = \Lam_r, \quad  
    \overline{M}_\P (\s=0) = M_\P^r .
      \label{renomalization condition at sigma=0}
\end{equation}
Equation (\ref{solution of tHooft-Weinberg equation}) is the general solution to the {}$^\pp$t Hooft-Weinberg equation (\ref{renormalization group equation 1}).

Replacing the momentum $e^\s k$ by $k$ and rewriting equation (\ref{solution of tHooft-Weinberg equation}), we obtain a slightly different form as
\begin{equation}
    \Gm_r^{(n)} \left( k/e^\s, t_r, \Lam_r, M_\P^{r 2},\mu \right) =
    \Gm_r^{(n)} \left( k,\overline{t}_r(\s), \overline{\Lam}(\s), \overline{M}_\P^2(\s),\mu \right) e^{-(4-nd_0)\s} .
         \label{tHooft-Weinberg equation 2}
\end{equation}
This expression is convenient to study the scaling behavior of the Green function with the momentum $k/e^\s$. Indicated by the physical momentum (\ref{physical momentum}), the scale parameter $\s$ can be identified with the constant mode of the $\phi$ field. Thus, the large $\s$ limit corresponds to the IR limit. Correspondingly, the limit $\s \to -\infty$ is the UV limit.

\subsection{Effective cosmological constant}
\noindent

As mentioned above, the dimensionless scale parameter $\s$ can be identified with the constant conformal mode. So, we introduce the constant background of the conformal mode by shifting the field as $\phi \to \phi+\s$.

Here, we consider the CFT limit of $t_r=0$. The large $b_1$ limit is also taken into account, while the ratios, $\Lam_r/b_1$ and $M_\P^{r 2}/b_1$, are taken to be the order of unity. In this limit, the one-loop approximation becomes valid and loop corrections to the effective action are written by functions of these ratios. The anomalous dimensions is then given by
\begin{equation}
     \gm_\Lam = \fr{4}{b_1} - \fr{18\pi^2}{b_1^2}\fr{M_\P^{r 4}}{\Lam_r}, \qquad 
     \gm_\EH = \fr{1}{b_1} 
      \label{anomalous dimension at large b_1}
\end{equation}
up to the order of $1/b_1$.

The effective cosmological constant is divided into three parts of tree, induced and loop terms. The tree term is $\Lam_r e^{4\s}$. The induced term is given by the finite term induced by the residue of the simple pole in (\ref{Laurent expansion of cosmological constant}). Since the residue $u_1$ is given by $\gm_\Lam$ in (\ref{anomalous dimension at large b_1}), we obtain 
\begin{equation}
       V^{\rm induced} = \left( 
                      \fr{4}{b_1}\Lam_r - \fr{18\pi^2}{b_1^2} M_\P^{r 4} \right) \s e^{4\s} .
\end{equation}

Since the expansion in the quantum field $\phi$ corresponds to the expansion in $1/b_1$, we expand the action up to the second order of the field. Rescaling the field to be $\phi=\sq{4\pi^2/b_1}\vphi$ in order to normalize the kinetic term of the conformal mode, we obtain
\begin{eqnarray}
    I \vert_{\vphi^2+{\rm c.t.}} &=& \int d^D x \Biggl\{ \half \vphi \pd^4 \vphi 
                + \fr{12\pi^2}{b_1} M_\P^{r 2} e^{2\s} \left( 2\vphi \pd^2 \vphi + \pd_\lam \vphi \pd_\lam \vphi \right)
             \nonumber \\
      && \qquad\quad
                + \fr{32\pi^2}{b_1}\Lam_r e^{4\s} \vphi^2 
                -\fr{1}{\bar{\eps}}\left( \fr{2}{b_1}\Lam_r - \fr{9\pi^2}{b_1^2}M_\P^{r 4} \right) e^{4\s}
                \Biggr\}. 
            \label{action up to second order}
\end{eqnarray}
Here, we add the counterterm to regularize the UV divergence. The dependence on arbitrary mass scale $\mu$ is suppressed for the present, because it can be easily recovered. The loop correction is then given by the one-loop diagram depicted in Fig.\ref{effective cosmological constant}.

\begin{figure}[h]
\begin{center}
\begin{picture}(200,100)(-20,10)
\Text(8,50)[]{$V^{\rm loop}=\sum$}
\CArc(100,50)(30,0,360) 
\Vertex(70,50){2}\Vertex(80,72){2}\Vertex(100,80){2}\Vertex(120,72){2}\Vertex(130,50){2}
\Vertex(80,28){2}\Vertex(100,20){2}\Vertex(120,28){2}
\Text(53,50)[]{$\Lam_r e^{4\s}$}\Text(64,76)[]{$\Lam_r e^{4\s}$}\Text(100,90)[]{$\Lam_r e^{4\s}$}\Text(140,76)[]{$\Lam_r e^{4\s}$}\Text(150,50)[]{$\Lam_r e^{4\s}$}
\Text(60,18)[]{$M_\P^{r 2} e^{2\s}$}\Text(100,8)[]{$M_\P^{r 2} e^{2\s}$}\Text(140,20)[]{$M_\P^{r 2} e^{2\s}$}
\end{picture} 
\end{center}
\caption{\label{effective cosmological constant}{\small Loop diagrams for the effective cosmological constant.}}
\end{figure}

We define the differential operator ${\cal D}$ such that the action is written in the form $\int \vphi {\cal D} \vphi/2$. In the momentum space, it is
\begin{equation}
    {\cal D} = k^4 - \fr{24\pi^2}{b_1} M_\P^{r 2} e^{2\s} k^2 + \fr{64\pi^2}{b_1}\Lam_r e^{4\s}.
\end{equation} 
The loop correction to the effective potential is then expressed as
\begin{eqnarray}
   V^{\rm loop} &=& - \log \left[ \det \left( {\cal D}_0^{-1} {\cal D} \right) \right]^{-1/2}
          \nonumber \\
       &=& \half \int \fr{d^D k}{(2\pi)^D} \log \left\{ 
            1 - \fr{24\pi^2}{b_1} M_\P^{r 2} e^{2\s} \fr{1}{k^2} + \fr{64\pi^2}{b_1} \Lam_r e^{4\s} \fr{1}{k^4}
            \right\} ,
\end{eqnarray}
where ${\cal D}_0=k^4$ is the inverse of the propagator of the rescaled conformal mode $\vphi$. Expanding the logarithmic function in series, we obtain the following expression:
\begin{eqnarray}
   V^{\rm loop} &=& \half \sum^\infty_{n=1} \fr{(-1)^{n-1}}{n} \int \fr{d^D k}{(2\pi)^D} 
           \left( \fr{64\pi^2}{b_1} \Lam_r e^{4\s} \fr{1}{k^4} - \fr{24\pi^2}{b_1} M_\P^{r 2}e^{2\s}\fr{1}{k^2}
               \right)^n
          \nonumber \\
      &=& \half \sum^\infty_{n=1} \fr{(-1)^{n-1}}{n} \sum^n_{m=0} \fr{n!}{(n-m)! m!} A^{n-m}(-B)^m I_{n;m} ,
           \label{series expansion}
\end{eqnarray}
where
\begin{equation}
      A= \fr{64\pi^2}{b_1} \Lam_r e^{4\s}, \qquad B = \fr{24\pi^2}{b_1}M_\P^{r 2} e^{2\s}
\end{equation}
and the momentum integral is defined by
\begin{equation}
     I_{n;m} = \int \fr{d^D k}{(2\pi)^D} \fr{1}{(k^2)^{2n-m}}
\end{equation} 
for $n \geq m$.

The momentum integral has the UV divergence for $2n-m = 2$ and the IR divergence for $2n-m \geq 2$, while vanishes for $2n-m <2$. The IR divergences are evaluated by introducing a small fictitious mass $z$, namely the square of momentum in the integrand is replaced as $k^2 \to k^2+z^2$. After carrying out the computation of the effective action, we take the vanishing limit of the mass $z$.

Since the integral $I_{1;1}$ vanishes at $z \to 0$, it gives no contribution. The integrals $I_{1;0}$ and $I_{2;2}$ have both the UV and IR divergences given by
\begin{equation}
    I_{1;0}=I_{2;2}= \fr{1}{(4\pi)^2} \left( \fr{1}{\bar{\eps}} - \log \fr{z^2}{\mu^2} \right).
\end{equation}
The integrals with $2n-m >2$ have the IR divergences as
\begin{equation}
    I_{n;m} = \fr{1}{(4\pi)^2} \fr{1}{(2n-m-1)(2n-m-2)} \left( \fr{1}{z^2} \right)^{2n-m-2}.
\end{equation}
Substituting these results into the expression (\ref{series expansion}) and subtracting the UV divergences by the counterterm in (\ref{action up to second order}), we obtain the following series: 
\begin{eqnarray}
     && V^{\rm loop} 
           \nonumber \\
     && = -\fr{A}{2(4\pi)^2} \log \fr{z^2}{\mu^2} + \fr{B^2}{4(4\pi)^2} \log \fr{z^2}{\mu^2}
           +\fr{AB}{4(4\pi)^2} \fr{1}{z^2} - \fr{A^2}{24(4\pi)^2} \fr{1}{z^4}
             \nonumber \\
     && \quad  
           + \fr{1}{(4\pi)^2} \sum^\infty_{n=3} \sum^n_{m=0} \fr{n!}{(n-m)! m!} 
                \fr{(-1)^{n-1}(-1)^m A^{n-m}B^m}{2n(2n-m-1)(2n-m-2)}
                 \left( \fr{1}{z^2} \right)^{2n-m-2} .
              \nonumber \\
       &&       \label{series for effective action}
\end{eqnarray}
The sum of the infinite series is evaluated and the expression of $V^{\rm loop}$ at $z \to 0$ is obtained in Appendix C. Adding the tree and induced terms, we finally obtain the following effective cosmological constant:
\begin{eqnarray}
    V &=& \Lam_r e^{4\s} +V^{\rm induced} + V^{\rm loop}
             \nonumber \\
     &=& e^{4\s} \Biggl\{  \Lam_r + 4 \s \left( 
                      \fr{\Lam_r}{b_1} - \fr{9\pi^2}{2}\fr{M_\P^{r 4}}{b_1^2} \right) 
             \nonumber \\
      && \qquad
            + \left( \fr{\Lam_r}{b_1} - \fr{9\pi^2}{2}\fr{M^{r 4}_\P}{b_1^2}  \right)
             \left[ 3 - \log \left(  \fr{(8\pi)^2}{\mu^4} \fr{\Lam_r}{b_1} e^{4\s}  \right) \right]
                   \nonumber \\
      && \qquad
            - 6\pi \fr{M^{r 2}_\P}{b_1} \sq{\fr{\Lam_r}{b_1}-\fr{9\pi^2}{4}\fr{M_\P^{r 4}}{b_1^2}} 
                   \arccos \left( \fr{3\pi}{2}\fr{M_\P^{r 2}/b_1}{\sq{\Lam_r/b_1}} \right) 
               \Biggr\} .
\end{eqnarray}
Here, note that the $\s$-dependences in the braces cancel out.

Let us consider the renormalization group improvement of the effective cosmological constant. By setting the conditions (\ref{renomalization condition at sigma=0}) and the anomalous dimensions (\ref{anomalous dimension at large b_1}), the renormalization group equations (\ref{running mass equation}) are solved as
\begin{eqnarray}
   \overline{\Lam}(\s) &=& e^{4\s} \left\{ \Lam_r + 4 \left( 
                      \fr{\Lam_r}{b_1} - \fr{9\pi^2}{2}\fr{M_\P^{r 4}}{b_1^2} \right) \s +\cdots \right\},
               \nonumber \\
    \overline{M}_\P^2(\s) &=& M_\P^{r 2} e^{2\s} \left( 1 + \fr{1}{b_1}\s + \cdots \right),
         \label{solution of running mass}
\end{eqnarray} 
and the effective cosmological constant satisfies the equation
\begin{equation}
     V\left( \overline{\Lam}(\s),\overline{M}_\P^2(\s),\mu \right) 
     =e^{4\s} V \left( \Lam_r, M_\P^{r 2}, \mu \right),
\end{equation}
where $V$ in the right-hand side is a constant evaluated at $\s=0$. Thus, the effective cosmological potential can be written in terms of the running mass scales as
\begin{eqnarray}
   V &=& \overline{\Lam}(\s) + \left( \fr{\overline{\Lam}(\s)}{b_1} 
          - \fr{9\pi^2}{2}\fr{\overline{M}^4_\P(\s)}{b_1^2}  \right)
            \left\{ 3 - \log \left(  \fr{(8\pi)^2}{\mu^4} \fr{\overline{\Lam}(\s)}{b_1}  \right) \right\}
                   \nonumber \\
     && - 6\pi \fr{\overline{M}^2_\P(\s)}{b_1} 
         \sq{\fr{\overline{\Lam}(\s)}{b_1}-\fr{9\pi^2}{4}\fr{\overline{M}_\P^4(\s)}{b_1^2}} 
             \arccos \left( \fr{3\pi}{2}\fr{\overline{M}_\P^2(\s)/b_1}{\sq{\overline{\Lam}(\s)/b_1}} \right) .
\end{eqnarray}

Both of the running mass scales decrease in the UV limit of $\s \to -\infty$ due to the exponential factor of $\s$.\footnote{ 
Since we consider the large $b_1$ expansion, the limit will be valid within $|\s/b_1| <1$. 
} 
In the IR limit of $\s \to \infty$, the running Planck mass monotonously increases. As for the running cosmological constant, however, if the correction term in the solution of $\Lam(\s)$ (\ref{solution of running mass}) is negative, it decreases even in the IR limit when this term becomes effective.


\section{Conclusion}
\setcounter{equation}{0}
\noindent

We studied the renormalizable quantum gravity formulated as a perturbed theory from CFT in four dimensions. The metric field was decomposed into the conformal mode and the traceless tensor mode, and the conformal mode was treated non-perturbatively without introducing its own coupling constant so that conformal symmetry was realized as a gauge symmetry, while the traceless tensor mode was handled in perturbation with the coupling constant indicating asymptotic freedom that measures a degree of deviation from CFT.

Dimensional regularization was used to carry out higher-order renormalization, which is a manifestly diffeomorphism invariant regularization at all orders. To determine the $D$-dimensional action, we applied the Wess-Zumino integrability condition to reduce indefiniteness existing in the four-derivative gravitational action. The renormalization was carried out by the counterterm method introducing the renormalization factor as usual, provided that the renormalization factor of the conformal mode is unity because of no coupling constant for this mode.

The effective action of quantum gravity improved by renormalization group was found. We then made clear the relationship among conformal anomalies, conformal symmetry and diffeomorphism invariance: the conformal anomaly can be divided into two groups of coupling-dependent and coupling-independent ones, and the former is ordinary conformal anomaly that violates conformal invariance, while the latter is, against its name, required for making conformal symmetry exact at the vanishing coupling limit. In any case, it was shown that conformal anomalies arise to guarantee diffeomorphism invariance quantum mechanically.

The anomalous scaling dimensions of the cosmological constant term and the Einstein action were calculated. We found that these results agree with those obtained by solving the physical state condition of conformal algebra in the CFT limit.

The renormalization group equation in which the scale parameter is identified with the constant conformal-mode was derived and applied it to the effective cosmological constant calculated in the large number limit of matter fields. We found that there is a solution that in the IR limit the the running Planck mass monotonously increases, while the running cosmological constant decreases, although both decrease in the UV limit as the perturbation theory is justified.

We managed the IR divergence by introducing a small fictitious mass for the gravitational field like a photon mass, because the lower-derivative gravitational action cannot be considered as a mass term due to the existence of the exponential factor of the conformal mode. We showed that the fictitious mass term is not gauge invariant and thus the IR divergence cancels out.

Recently, it has been demonstrated that the conformal symmetry mixes the positive-metric and negative-metric modes, and consequently the negative-metric mode does not appear independently as a gauge invariant state at all \cite{hamada08,hh,hamada05}. Thus, the physical state is restricted to be a diffeomorphism invariant combination of these modes, classified by real fields with even number of derivative such as the scalar curvature. The correctness of the overall sign of the Riegert and Weyl actions, not the sign of each mode, will be significant to make two-point correlation functions of these real fields positive.

The conformal invariance forces us change the aspect of space-time at very short scales less than the Planck length, where there is no particle picture propagating on the flat background. Hence, a traditional S-matrix description is not adequate at all. The fact that there is no $\hbar$ in the Weyl action and the Wess-Zumino action also indicates this description. Thus, it is expected that unphysical states are confined as a virtual quantum state and does not appear in the classical limit $\hbar \to 0$.

\vspace{5mm}


\appendix

\vspace{5mm}
\begin{center}
   {\Large {\bf APPENDIX}}
\end{center}

\section{Various Formulae for Gravitational Fields}
\setcounter{equation}{0}
\noindent

Our curvature conventions are given by $R^{\lam}_{~\mu\s\nu} = \pd_{\s}\Gm^{\lam}_{~\mu\nu} +\cdots$ and $R_{\mu\nu} = R^{\lam}_{~\mu\lam\nu}$ such that the commutator of the covariant derivatives satisfies 
\begin{equation}
   \left[ \nb_\mu, \nb_\nu \right] A_{\lam_1,\cdots \lam_n}
     = \sum_{i=1}^n R^{~~~~\s_i}_{\mu\nu\lam_i} A_{\lam_1, \cdots, \s_i, \cdots, \lam_n}  .
\end{equation}

\paragraph{Conformal variations in $D$ dimensions}
Under the Weyl rescaling $\dl_{\om}g_{\mu\nu} =2\om g_{\mu\nu}$, the scalar curvature transforms as
\begin{equation}
  \dl_{\om} \sq{g}R = (D-2)\om \sq{g}R -2(D-1)\sq{g}\nb^2 \om ,
\end{equation}
and the fourth-order gravitational quantities transform as follows:
\begin{eqnarray}
   \dl_{\om} \sq{g}R^{\mu\nu\lam\s}R_{\mu\nu\lam\s} 
     &=& (D-4) \om \sq{g}R^{\mu\nu\lam\s}R_{\mu\nu\lam\s} 
      -8\sq{g} R^{\mu\nu}\nabla_{\mu}\nb_{\nu} \om , 
          \nonumber \\ 
   \dl_{\om} \sq{g}R^{\mu\nu}R_{\mu\nu} 
     &=& (D-4) \om \sq{g}R^{\mu\nu}R_{\mu\nu} 
         -2 \sq{g}R \nb^2 \om 
           \nonumber  \\ 
     && 
      -2(D-2)\sq{g} R^{\mu\nu}\nb_{\mu}\nb_{\nu} \om ,    
           \nonumber \\ 
   \dl_{\om} \sq{g}R^2
     &=& (D-4) \om \sq{g}R^2 - 4(D-1) \sq{g}R \nb^2 \om , 
           \nonumber \\ 
   \dl_{\om} \sq{g}\nb^2 \!R 
     &=& (D-4)\om \sq{g} \nb^2 \!R 
         +(D-6)\sq{g} \nb^{\lam}R \nb_{\lam}\om 
             \nonumber \\ 
     && 
         -2 \sq{g}R \nb^2 \om -2(D-1) \sq{g}\nb^4 \!\om . 
\end{eqnarray}

\paragraph{Mode expansions}
The metric field is decomposed into the conformal mode $\phi$ and the traceless tensor modes $h_{\mu\nu}$ as $g_{\mu\nu}=e^{2\phi}\bg_{\mu\nu}$ with $\bg_{\mu\nu}=(\hg e^h)_{\mu\nu}$, where we suppress the coupling constant $t$. The gravitational quantities are then decomposed as
\begin{eqnarray}
   \Gm^{\lam}_{~\mu\nu} 
    &=& {\bar \Gm}^{\lam}_{~\mu\nu} +\bg^{\lam}_{~\mu}\bnb_{\nu}\phi 
       +\bg^{\lam}_{~\nu}\bnb_{\mu}\phi -\bg_{\mu\nu} \bnb^{\lam}\phi,
           \nonumber \\
   R &=& \e^{-2\phi} \left\{ \bR -2(D-1)\bnb^2 \phi 
        -(D-1)(D-2)\bnb_{\lam}\phi \bnb^{\lam}\phi \right\} ,
             \nonumber \\
    R_{\mu\nu} 
    &=&\bR_{\mu\nu}-(D-2)\bDelta_{\mu\nu} 
        -\bg_{\mu\nu} \left\{ \bnb^2 \phi 
        +(D-2)\bnb_{\lam}\phi \bnb^{\lam}\phi \right\},
               \nonumber  \\
   R^{\lam}_{~\mu\s\nu}
   &=& \bR^{\lam}_{~\mu\s\nu} + \bg^{\lam}_{~\nu}\bDelta_{\mu\s}   
       -\bg^{\lam}_{~\s}\bDelta_{\mu\nu}+\bg_{\mu\s}\bDelta^{\lam}_{~\nu}
       -\bg_{\mu\nu}\bDelta^{\lam}_{~\s} 
               \nonumber  \\
   &&  + \bigl( \bg^{\lam}_{~\nu}\bg_{\mu\s}
            -\bg^{\lam}_{~\s}\bg_{\mu\nu} \bigr) 
               \bnb_{\rho}\phi\bnb^{\rho}\phi, 
\end{eqnarray}
where $\bDelta_{\mu\nu}=\bnb_{\mu}\bnb_{\nu}\phi-\bnb_{\mu}\phi\bnb_{\nu}\phi$. The quantities with the bar are expanded in the traceless tensor mode as
\begin{eqnarray}
   \bar{\Gm}^\lam_{\mu\nu} &=& \hat{\Gm}^\lam_{\mu\nu}
         + \hnb_{(\mu}h^\lam_{~\nu)} - \half \hnb^\lam h_{\mu\nu}
         + \half \hnb_{(\mu} (h^2)^\lam_{~\nu)} - \fr{1}{4} \hnb^\lam (h^2)_{\mu\nu}
              \nonumber \\
         &&
           - h^\lam_{~\s} \hnb_{(\mu} h^\s_{~\nu)} + \half h^\lam_{~\s}\hnb^\s h_{\mu\nu} 
           +o(h^3),
              \nonumber \\
   \bR &=&  \hR -\hR_{\mu\nu} h^{\mu\nu} + \hnb_\mu \hnb_\nu h^{\mu\nu} 
         - \fr{1}{4} \hnb^\lam h^\mu_{~\nu} \hnb_\lam h^\nu_{~\mu}
          + \half \hR^\s_{~\mu\lam\nu} h^\lam_{~\s} h^{\mu\nu} 
                \nonumber \\
       && 
          + \half \hnb_\nu h^\nu_{~\mu} \hnb_\lam h^{\lam\mu}
          -\hnb_\mu ( h^\mu_{~\nu} \hnb^\lam h^\nu_{~\lam}) + o(h^3),
                \nonumber \\
   \bR_{\mu\nu} &=& \hR_{\mu\nu}  -\hR^\s_{~\mu\lam\nu} h^\lam_{~\s}
                    + \hR^\lam_{(\mu}h_{\nu)\lam} + \hnb_{(\mu} \hnb^\lam h_{\nu)\lam} 
                    -\half \hnb^2 h_{\mu\nu}
                \nonumber \\
      && 
         -\half h^\lam_{(\mu} \hnb^2 h_{\nu)\lam} 
         -\half \hnb^\lam h^\s_{~\mu} \hnb_\s h_{\nu\lam}
         -\fr{1}{4} \hnb_\mu h^\lam_{~\s} \hnb_\nu h^\s_{~\lam}
                \nonumber \\
      &&
         -\half \hnb_\lam ( h^\lam_{~\s} \hnb_{(\mu} h^\s_{~\nu)})
         +\half \hnb_\lam ( h^\s_{(\mu} \hnb_{\nu)} h^\lam_{~\s})
         +\half \hnb_\lam ( h^\lam_{~\s} \hnb^\s h_{\mu\nu}) + o(h^3) ,
                \nonumber \\
\end{eqnarray}
where $\bR=\bg^{\mu\nu}\bR_{\mu\nu}$.

\section{Useful Formulae for Dimensional Regularization}
\setcounter{equation}{0}
\noindent

We here summarize various formulae to evaluate the $D$-dimensional integrals in dimensional regularization. 

\paragraph{Fundamental integral formulae}
The volume integral in $D$ dimensional Euclidean momentum space is given by
\begin{eqnarray}
    \int d^D k &=& \int k^{D-1}dk \int d\Om_D, \quad (k^2=k_\mu k_\mu )
          \nonumber \\
    \int d\Om_D &=& \int \prod^{D-1}_{l=1}  \sin^{D-1-l}\theta_l d\theta_l
                = \fr{2\pi^{D/2}}{\Gm \left( \fr{D}{2} \right)} .
\end{eqnarray}
The following integral formulae are useful:
\begin{equation}
     \int \fr{d^D k}{(2\pi)^D} \fr{k^{2n}}{(k^2+L)^\a}
      = \fr{1}{(4\pi)^{D/2}} 
          \fr{\Gm\left(n+\fr{D}{2}\right)\Gm\left(\a-n-\fr{D}{2}\right)}
             {\Gm\left(\fr{D}{2}\right)\Gm(\a)}
               L^{D/2+n-\a} .
               \label{fundamental integral}
\end{equation}
and 
\begin{eqnarray}
     \int \fr{d^D k}{(2\pi)^D} k_\mu k_\nu f(k^2)
       &=& \fr{1}{D}\dl_{\mu\nu} \int \fr{d^D k}{(2\pi)^D} k^2 f(k^2),
             \nonumber \\
     \int \fr{d^D k}{(2\pi)^D} k_\mu k_\nu k_\lam k_\s f(k^2)
      &=& \fr{1}{D(D+2)} \left( \dl_{\mu\nu}\dl_{\lam\s} + \dl_{\mu\lam}\dl_{\nu\s}
                   + \dl_{\mu\s}\dl_{\nu\lam} \right)
             \nonumber \\
       && \qquad\qquad \times
               \int \fr{d^D k}{(2\pi)^D} k^4 f(k^2) .
\end{eqnarray}
Here, the integral with odd number of $k_\mu$ vanishes.

\paragraph{Feynman parameterization}
In order to evaluate more complicated integrals that appear in self-energy diagrams and so on, the Feynman parameterization is often used,
\begin{equation}
    \fr{1}{A^\a B^\b} = \fr{\Gm(\a+\b)}{\Gm(\a)\Gm(\b)}
          \int^1_0 dx \fr{(1-x)^{\a-1}x^{\b-1}}{[(1-x)A+xB]^{\a+\b}} .
\end{equation}
Applying this formula to the self-energy integral with $A=k^2+z^2$ and $B=(k+l)^2+z^2$, the integral can reduce to the fundamental form (\ref{fundamental integral}) as follows:
\begin{eqnarray}
   &&\int \fr{d^D k}{(2\pi)^D} \fr{f(k_\mu,l_\nu)}{(k^2 +z^2)^\a ((k+l)^2+z^2)^\b}
          \nonumber \\
   && = \fr{\Gm(\a+\b)}{\Gm(\a)\Gm(\b)}
          \int^1_0 dx (1-x)^{\a-1}x^{\b-1} \int \fr{d^D k^\pp}{(2\pi)^D}
          \fr{f(k_\mu^\pp-xl_\mu,l_\nu)}{[ k^{\pp 2}+z^2 +x(1-x)l^2 ]^{\a+\b}} .
          \nonumber \\
   &&
\end{eqnarray}

\paragraph{Extraction of UV divergences}
The UV divergence arises as a pole of $\eps=(4-D)/2$. To extract the pole, the following formulae are useful: 
\begin{eqnarray}
     \Gm(\eps) &=& \fr{1}{\eps} -\gm 
          + \fr{\eps}{2} \left( \gm^2 + \fr{\pi^2}{6} \right) +o(\eps^2),
        \nonumber \\
      a^\eps &=& e^{\eps \ln a}= 1+ \eps \ln a +o(\eps^2).
\end{eqnarray}
Here, $\gm=0.57721\ldots$ is the Euler constant and $a$ represents the quantities of $k^2$ and $z^2$, for instance.

\paragraph{Dirac gamma matrices}
The Dirac gamma matrices in $D$ dimensions satisfy the following relations:
\begin{eqnarray}
   \gm_\lam \gm_\lam &=& -D, 
      \nonumber \\
   \gm_\lam \gm_\mu \gm_\lam &=&(D-2)\gm_\mu,
      \nonumber \\
   \gm_\lam \gm_\mu \gm_\nu \gm_\lam &=& -(D-4)\gm_\mu \gm_\nu + 4\dl_{\mu\nu}.
\end{eqnarray}

\section{Evaluation of The (\ref{series for effective action}) Series}
\setcounter{equation}{0}
\noindent

Introducing new variables
\begin{equation}
    x^2 = \fr{A}{z^4},  \qquad xy = \fr{B}{z^2},
\end{equation}
we rewrite the sum of infinite series in (\ref{series for effective action}), which is denoted by $z^4 f(x,y)/(4\pi)^2$ with 
\begin{equation}
    f(x,y) = \sum_{n=3}^\infty \sum^n_{m=0} \fr{n!}{(n-m)! m!} 
                \fr{(-1)^{n-1}(-1)^m}{2n(2n-m-1)(2n-m-2)} x^{2n-m} y^m .
\end{equation}

Consider the function given by differentiating $f(x,y)/x$ twice with respect to $x$. It can be evaluated as
\begin{eqnarray}
    h(x,y) &=& \fr{\pd^2}{\pd x^2} \left\{ \fr{1}{x}f(x,y) \right\} 
          \nonumber \\
    &=& \sum_{n=3}^\infty \fr{(-1)^{n-1}}{2n} \sum^n_{m=0} \fr{n!}{(n-m)! m!} x^{2n-m-3} (-y)^m 
          \nonumber \\
    &=& \sum_{n=3}^\infty \fr{(-1)^{n-1}}{2n} \fr{1}{x^3} (x^2 -xy)^n
          \nonumber \\
    &=& \fr{1}{2x^3} \left\{ \log \left( 1+x^2-xy \right) -x^2 +xy + \half \left( x^2 -xy \right)^2 \right\} .
\end{eqnarray}
Integrating two times with respect to $x$, we obtain the following result:
\begin{eqnarray}
    f(x,y) &=& x \int^x_0 du \int^u_0 dv h(v,y)
                  \nonumber \\
           &=& \left\{ \fr{1}{4} \left( 1-x^2 \right) -\fr{1}{4}xy + \fr{1}{8}x^2 y^2 
                          \right\} \log \left( 1+x^2-xy \right)
                   \nonumber \\
           &&  + \fr{3}{4}x^2 +\fr{1}{24}x^4 + \fr{1}{4} \left( x-x^3 \right)y -\fr{3}{8} x^2 y^2
                  \nonumber \\
           &&  - \fr{1}{4\sq{4-y^2}} \left( 8x -4x^2 y -2x y^2 +x^2 y^3 \right) 
                    \arctan \left( \fr{2x\sq{4-y^2}}{4-2xy} \right)  .
                  \nonumber \\ 
           &&        
\end{eqnarray}

Using this expression, we can obtain the loop correction to the effective potential by taking the vanishing limit of the mass scale $z$ as
\begin{eqnarray}
   V^{\rm loop} &=& \lim_{z\to 0} \fr{1}{(4\pi)^2} \Biggl\{ 
           -\fr{A}{2} \log \fr{z^2}{\mu^2} + \fr{B^2}{4} \log \fr{z^2}{\mu^2}
           +\fr{AB}{4} \fr{1}{z^2} - \fr{A^2}{24} \fr{1}{z^4}
                 \nonumber \\
          && \qquad\qquad\quad
             + z^4 f \left( \fr{\sq{A}}{z^2}, \fr{B}{\sq{A}} \right) \Biggr\}
                 \nonumber \\
          &=& \fr{1}{(4\pi)^2} \left\{ \fr{1}{8} \left( 2A-B^2 \right) 
                                \left( 3-\log \fr{A}{\mu^4} \right) 
                     - \fr{B}{4} \sq{4A-B^2} \arccos \left( \fr{B}{2\sq{A}} \right)  \right\} .
                 \nonumber \\
          &&
\end{eqnarray}
Here, we assume $B < 2\sq{A}$ for the present and use the formula: $\arctan(\sq{1-w^2}/w)=\arccos w$. The IR divergences at $z = 0$ indeed cancel out. This result can be extended to the range of $B > 2\sq{A}$ using the expression of the arccos function: $\arccos w = i\log(w+\sq{w^2-1})$ with $w >1$. If we take the limit of $A \to 0$, $V^{\rm loop}$ reduces to $B^2\{-3+\log (B^2/\mu^4)\}/8(4\pi)^2$.


\end{document}